\documentclass[% 
reprint,
showpacs,
amsmath, amssymb,
aps,
pre]{revtex4-1}
\usepackage[hang, nooneline, FIGBOTCAP]{subfigure}

\usepackage{graphicx}% 
\usepackage{dcolumn}% 
\usepackage{bm}
\usepackage{amsmath}% 
\newcommand{\bs}[1] {\boldsymbol{#1}}
\usepackage{xspace}
\newcommand{\eg}{e.g.,\xspace}
\newcommand{\ie}{i.e.,\xspace}
\newcommand{\etal}{\emph{et al.}\xspace}
\newcommand{\etc}{etc.\xspace}
\newcommand{\ignore}[1]{}

\newcommand{\physicaa}{Physica A}
\newcommand{\jphysamg}{J. Phys. A -- Math. Gen.}

\newcommand{\jphysc}{J. Phys. C -- Solid State Phys.}
\newcommand{\jstatphys}{J. Stat. Phys.}

\newcommand{\physrep}{Phys. Rep.}

\newcommand{\epl}{Europhys. Lett.}

\newcommand{\communmathphys}{Commun. Math. Phys.}

\newcommand{\jstatmech}{J. Stat. Mech.}
\newcommand{\advphys}{Adv. Phys.}

\newcommand{\jphyscondmat}{J. Phys. Cond. Mat.}

\newcommand{\jrsocinterface}{J. R. Soc. Interface}
\newcommand{\physbiol}{Phys. Biol.}

\newcommand{\jmathbiol}{J. Math. Biol.}

\newcommand{\pnas}{Proc. Natl. Acad. Sci. USA}
\newcommand{\progtheorphys}{Prog. Theor. Phys.}
\newcommand{\jtheorbiol}{J. Theor. Biol.}

\begin{document}

\title{Exclusion processes: short range correlations induced by adhesion 
and contact interactions}

\author{Gianluca Ascolani}
\email{ascolani@imnc.in2p3.fr}
\affiliation{CNRS, UMR 8165, IMNC, Univ Paris-Sud, Univ Paris Diderot, 
F-91405 Orsay, France}

\author{Mathilde Badoual}
\email{badoual@imnc.in2p3.fr}
\affiliation{Univ Paris Diderot, Laboratoire IMNC, UMR 8165 CNRS, Univ 
Paris-Sud, F-91405 Orsay, France}

\author{Christophe Deroulers}
\email{deroulers@imnc.in2p3.fr}
\affiliation{Univ Paris Diderot, Laboratoire IMNC, UMR 8165 CNRS, Univ 
Paris-Sud, F-91405 Orsay, France}

\date{5 October 2012}% 

\begin{abstract}
We analyze the out-of-equilibrium behavior of exclusion processes where agents interact with their nearest neighbors, and we study the short-range correlations which develop because of the exclusion and other contact interactions. 
The form of interactions we focus on, including adhesion and contact-preserving interactions, is especially relevant for migration processes of living cells. 
We show the local agent density and nearest-neighbor two-point correlations resulting from simulations on two dimensional lattices in the transient regime where agents invade an initially empty space from a source and in the stationary regime between a source and a sink. 
We compare the results of simulations with the corresponding quantities derived from the master equation of the exclusion processes, 
and in both cases, we show that, during the invasion of space by agents, a wave of correlations travels with velocity $v(t) \sim t^{-1/2}$.
The relative placement of this wave to the agent density front and the time dependence of its height may be used to discriminate between different forms of contact interactions or to quantitatively estimate the intensity of interactions. 
We discuss, in the stationary density profile between a full and an empty reservoir of agents, the presence of a discontinuity close to the empty reservoir. 
Then, we develop a method for deriving approximate hydrodynamic limits of the processes. From the resulting systems of partial differential equations, we recover 
the self-similar behavior of the agent density and correlations during space invasion.

\end{abstract}

\maketitle

\section{\label{sec:level1}Introduction}

 Since their introduction in the 1940s~\cite{von-neumann-livre-automates-cellulaires}, cellular automata 
have become an essential tool to study collective behavior in complex systems starting from the individual level in many 
areas of science: fluid dynamics~\cite{frisch-hasslacher-pomeau-lattice-gas-navier-stokes, 
dhumieres-lallemand-frisch-lattice-gas-3d, rothman-zaleski-revue-gaz-sur-reseau-avec-limite-hydrodynamique}, 
reaction-diffusion problems~\cite{boon-et-al-revue-lattice-gas-avec-reaction}, dynamics of 
glasses~\cite{ritort-sollich-revue-modeles-a-contraintes-cinetiques}, 
epidemiology~\cite{smith-et-al-modele-spatial-pour-la-rage}, traffic 
flow~\cite{chowdhury-santen-schadschneider-revue-trafic-vehicules, 
helbing-revue-trafic, hilhorst-appert-multi-lane-tasep-pour-pietons}. In 
the recent years, there have been many applications in biology, both at 
the intracellular and tissue levels~\cite{deutsch-dormann-livre, 
anderson-chaplain-rejniak-livre, 
alber-kiskowski-glazier-jiang-revue-automates-cellulaires}.

Special cases of cellular automata are the exclusion processes  which have been successfully applied to study motility problems when the concentration of agents is such that the geometric hindrance which they impose on each other cannot be neglected~\cite{liggett-livre-1985, liggett-livre-1999}.
In the exclusion processes, each lattice site is occupied by at most one agent, so that steric effects (hard-core repulsion) are incorporated from the very beginning and lead to nontrivial effects even in absence of other interactions~\cite{spohn-livre-limite-hydrodynamique, appert-derrida-lecomte-vanwijland-courant-ssep}. 
Ulterior interactions between agents may be added to study intermolecular~\cite{garrahan-sollich-toninelli-revue-modeles-a-contraintes-cinetiques}, intercellular or interindividual  relationship~\cite{helbing-revue-trafic, simpson-et-al-catalogue-interactions-de-contact}.

In practice, these 
interactions are specified using \emph{rules}, that is, expressions for 
the time rate of jump of each agent to another (empty) lattice site as a 
function of the present content of each lattice site. All what is needed 
to define the model is the lattice geometry, not necessarily regular, and the list of rules.

 Although cellular automata were designed to be efficiently simulated on a 
computer, it is helpful to supplement them with a macroscopic 
description of the collective behavior of agents, which often takes the 
form of a partial differential equation (PDE). Even in cases where 
simulations are available (usually repeated a large number of times to 
take into account stochastic noise ~\cite{liggett-livre-1999, 
schmeiser-anguige-limite-continue-1d-avec-adhesion}), the PDE may be a compact way to 
give an overview of the large scale behavior of the system, to distinguish the 
universal features of a family of related exclusion processes in the 
spirit of a RG-like approach~\cite{taueber-et-al-revue-rg-pour-reaction-diffusion}, or to 
classify them~\cite{simpson-et-al-catalogue-interactions-de-contact, 
murray-et-al-classification-forces-via-limite-continue}. 

If one wants to study the effects of variations of the model parameters, solving PDEs is faster than performing stochastic simulations,  and  PDEs are also useful to retrieve analytic results. 
In cases where simulations of the exclusion process is intractable because the number of agents in a 
realistic system is too large (the human body contains~$\simeq 10^{14}$ 
cells, the human brain~$\simeq 3.10^{11}$), or because the exclusion 
process has to be embedded in a larger system in a multiscale approach, 
this macroscopic description is essential.

 The proof of existence of a PDE describing an exclusion process has 
been the subject of quite involved mathematical 
developments~\cite{de-masi-presutti-livre-limite-hydrodynamique, 
kipnis-landim-livre-limite-hydrodynamique, 
fritz-preuve-diffusion-non-lineaire-a-partir-de-limite-hydro, 
guo-papanicolaou-varadhan-preuve-diffusion-non-lineaire-a-partir-de-limite-hydro, 
varadhan-preuve-diffusion-non-lineaire-a-partir-de-limite-hydro, 
varadhan-yau-preuve-diffusion-non-lineaire-a-partir-de-limite-hydro, 
goncalves-landim-toninelli-porous-medium-equation-comme-limite-hydrodynamique}. 
Usually, to get an explicit expression for the PDE, one uses a simple 
approximate technique based on the Chapman-Enskog 
expansion~\cite{chapman-cowling-livre}. First, a system of coupled 
ordinary differential equations (ODEs) for the average number of agents 
in each site of the lattice, or equivalently the occupation probability 
or density of each site of the lattice, is derived from the rules which 
define the exclusion process (possibly through the use of the so-called 
master equation~\cite{vankampen-livre}). Because of agent interactions 
and exclusion, each equation generally involves joint probabilities or 
correlation functions, like the probability that two 
nearest-neighbor sites are both occupied at the same time. Then, a 
mean-field-like approximation is made to express these correlation 
functions as product of site occupation probabilities, as if the 
occupations of two sites were statistically independent.

 Finally, assuming that the occupation probability of each site is a regular 
function of the position when the lattice step tends to zero, 
which amounts to say that the typical length scale over which this 
probability varies is much longer than one lattice step, a Taylor 
expansion of this function is substituted into the system of ODEs. Truncating 
the result to lowest non-vanishing order, one is left with a PDE for the 
density of agents as a function of space and time. For completeness, let 
us mention that such a derivation of a macroscopic model from a 
discrete, microscopic model can be done in many other settings, \eg to 
quote a few that were used in biological modeling, the cellular Potts 
model~\cite{painter-discrete-continuous-cell-movement, 
alber-et-al-limite-continue-potts-1, 
alber-et-al-limite-continue-potts-2}, cellular automata on a disordered 
lattice~\cite{drasdo-coarse-graining}, lattice-gas cellular 
automata~\cite{hatzikirou-brusch-deutsch-limite-continue-lgca}, and discrete 
models with forces~\cite{murray-et-al-limite-continue-1d-avec-forces}.

 In the case of exclusion processes where agents can only jump to 
nearest neighboring sites and where the rules involve only a 
short-range interaction between 
agents~\cite{drasdo-et-al-classification-croissance-prl}, the PDE often 
takes the form of a nonlinear diffusion equation, and the diffusivity 
depends on the local 
density~\cite{guo-papanicolaou-varadhan-preuve-diffusion-non-lineaire-a-partir-de-limite-hydro, 
varadhan-preuve-diffusion-non-lineaire-a-partir-de-limite-hydro, 
varadhan-yau-preuve-diffusion-non-lineaire-a-partir-de-limite-hydro, 
boon-et-al-limite-hydro-diffusion-non-lineaire-epl, 
boon-et-al-limite-hydro-diffusion-non-lineaire-pre, 
boon-et-al-limite-hydro-diffusion-non-lineaire-morphogene}. The 
interested reader will find explicit expressions of the nonlinear 
diffusivity for a large number of such exclusion processes in two recent 
works by Fernando 
\etal~\cite{simpson-et-al-catalogue-interactions-de-contact} and by 
Penington \etal~\cite{penington-et-al-generalisation-limite-continue}.

 This simple, mean-field-like approximation works remarkably well when 
the large-scale behavior of the system is of diffusive nature, \ie when 
the nonlinear diffusion coefficient is positive for all local densities 
of agents. On the contrary, a negative value of the diffusivity is the 
sign that the microscopic dynamics tend to form aggregates or is 
subject to demixion. In that case, the average occupation number of 
sites varies on the length scale of one or a few lattice steps, hence 
the hypothesis of regularity of the density as a function of position 
when the lattice step vanishes is inconsistent, this function cannot satisfy a 
PDE, and the approach breaks down.

 However, even when the diffusivity is always positive, the agreement 
between the density of agents predicted from the PDE and the density obtained 
through an average over many simulations of the exclusion process is not 
perfect~\cite{limite-continue-modele-ma-champ-moyen, 
simpson-et-al-adhesion}, and this has been proven to be due to 
correlations between the occupations of neighboring 
sites~\cite{limite-continue-modele-ma-champ-moyen}. Usually, the next 
step beyond the mean-field approximation is the so-called pair 
approximation, when one keeps track, at the macroscopic level, of both 
the probability of occupation of each site and of joint probabilities 
that two sites are occupied at the same time (hereafter called two-point 
correlation functions).
This has been used, in the case of macroscopically spatially uniform systems, in 
condensed matter 
physics~\cite{butcher-summerfield-pair-approximation-hopping-conductivity}, 
to study random walks~\cite{tahirkheli-eliott-diffusion-traceur}, 
reaction-diffusion problems~\cite{rudavets-pairing-correlations}, 
epidemic models~\cite{joo-lebowitz-pair-approximation-dans-sirs}, 
ecology~\cite{matsuda-et-al-lotka-volterra-avec-pair-approximation, 
ellner-pair-approximation-with-multiple-interaction-scales}, and
multicellular 
systems~\cite{simpson-baker-correcting-mean-field-uniform-in-space}. 
Recently, Simpson and Baker extended that approach to one-dimensional 
systems that are macroscopically non uniform, as during an invasion 
process. They found an excellent agreement of the macroscopic model with 
stochastic simulations using two-point correlation functions of site 
distances by up to two lattice 
steps~\cite{simpson-baker-correcting-mean-field-adhesion-1d} or ten lattice 
steps~\cite{baker-simpson-correcting-mean-field-spatially-dependent}.

 In this work, we use a similar approach to study spatially non-uniform 
systems in two dimensions, keeping the macroscopic model simple by using 
only nearest-neighbor two-point correlation functions, and casting it 
in the form of coupled PDE. While the model is simple, its results agree 
much better with stochastic simulations than a PDE for the density 
alone. Its mathematical expression as PDE allows us to analyze the 
self-similarity (or scaling) properties of its solutions in the context 
of invasion of space from a source of cells, as in 
wound-healing-like~\cite{khain-et-al-hypoxie} or migration 
assay~\cite{aubert-et-al-migration-sans-astrocytes-sains} experiments. 
We show that the scaling properties of correlations may help to 
distinguish between several microscopic mechanisms not only in theory, but also in experiments.

We have in mind applications to living cell migration processes, which are essential in a number of biological contexts like development, repair, tumor and cancer progression; therefore, we restrict ourselves to the family of exclusion processes where
the rate of movement of one cell depends only on the present contacts before moving (which may be preserved or lost) and not on the future contacts (contacts with cells which will be nearest-neighbor only after the move)  --- 
``direction then interactions'' in the terminology 
of~\cite{simpson-et-al-catalogue-interactions-de-contact}. This is a 
realistic setting to study contact interactions 
(adhesion~\cite{schmeiser-anguige-limite-continue-1d-avec-adhesion, 
khain-et-al-cell-cell-adhesion-wound-healing, 
khain-et-al-adhesion-lignees-de-gliomes-sans-migration, 
khain-et-al-hypoxie, simpson-et-al-adhesion} and cell-cell communication 
phenomena~\cite{aubert-et-al-migration-sans-astrocytes-sains, 
aubert-et-al-migration-avec-astrocytes-sains, 
limite-continue-modele-ma-champ-moyen, 
mesnil-et-al-gap-junctions-cancer}), disregarding, \eg chemotaxis or 
quorum sensing. But the method can be extended to more general exclusion 
processes. A potential application is personalized treatments of 
invasive tumors such as 
glioma~\cite{effet-inhibition-jonctions-communicantes}, where computer 
simulations of a mathematical model fed by patient-specific parameters 
will help providing the best therapeutic strategy, guide surgical 
resection, radiotherapy or chemoterapy, and so on. There, giving an 
accurate prediction of the amount of infiltrated cells in each part of 
the brain will be essential.

 This family of exclusion processes is introduced in 
Section~\ref{sec:level2}. In Section~\ref{sec:level3}, we derive the 
usual mean-field macroscopic approximation of them, as well as our 
improved macroscopic models. We compare them to stochastic simulations 
of the exclusion processes. In Section~\ref{sec:level4}, we go to the 
continuous space limit and study the self-similarity behavior of the 
solutions. Finally, we give a discussion and conclusions in 
Section~\ref{sec:level5}.

\section{\label{sec:level2}The model}

 In our exclusion processes \cite{liggett-livre-1999, baker-rep}, agents move on a fixed $d$-dimensional 
lattice, each site of which may contain 0 or 1 agent. For simplicity, we 
will apply our results only to the bidimensional hexagonal tiling 
(triangular lattice), but they can be extended straightforwardly. In a 
move, an agent can only jump to an empty nearest neighbor site. Of 
course, the reality of biological movement is much more complicated than 
this (for instance, cells deform, make protrusions \etc), but the aim is 
to gain access to the macroscopic collective behavior for which we 
believe that a too detailed description may be irrelevant because many 
microscopic degrees of freedom will be ``forgotten'' at large scales 
and will make simulations and computations very 
difficult if possible at all.

\subsection{Jump rates and interactions}

 The definition of a process is completed with the specification of the 
rate (probability of occurrence per time unit) of each jump of an agent, which is 
assumed not to depend explicitly on time or position, but only of the 
content of the lattice sites.

\begin{figure}[h]
\includegraphics[width=0.45\textwidth, height=0.5\textwidth, angle=-90]{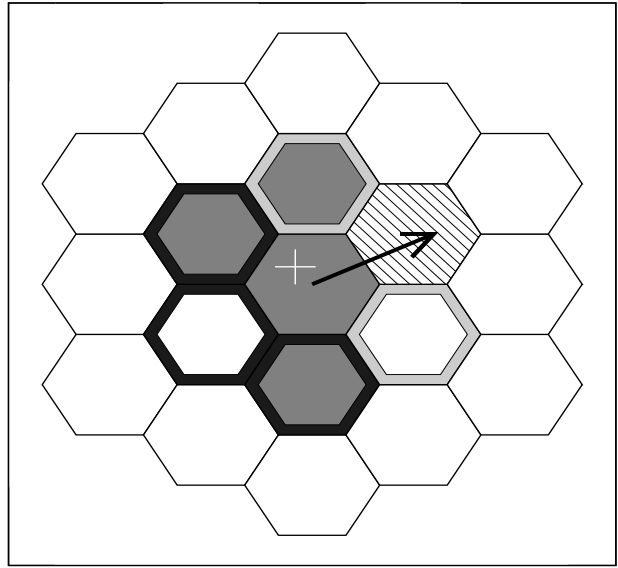}
\caption{\label{LatRule}Analytical computations and numerical simulations are done in a hexagonal tiling.
The distance between two nearest neighbor sites is $\lambda$. 
Gray filled sites are occupied by cells; empty sites are filled in white. The cell marked with a white cross symbol attempts to move in the direction of the arrow, if the hatched site is empty. 
The neighbor sites $\mathcal{N}$  and $\mathcal{M}$ of the marked cell have black and light gray border respectively. 
Cells in the sites with a black border form non-maintained links, which break during the jump of the marked cell. Cells in the common neighbor sites between the marked cell and the hatched site form maintained links preserved during the jump of the marked cell.}
\end{figure}

 Let $i$ and $j$ be two lattice sites. We denote $\mathcal{V}(i)$ --- 
for \emph{vicinity} --- the set of the nearest-neighbors of $i$ on the 
lattice, $V(i)$ the numbers of these sites, and $v(i)$ the number of 
sites among them which are occupied ($0 \le v(i) \le V(i)$, $V(i)=V=6$ on the 
hexagonal tiling; $v(i)$ may vary with time, but not $V(i)$). Likewise, 
$\mathcal{M}(i, j)$ is the set of lattice sites which are nearest 
neighbors of both $i$ and $j$ (but distinct from $i$ and $j$) --- 
$\mathcal{M}$ for \emph{maintained} contact ---, $M(i, j)$ their numbers 
(2 on the hexagonal tiling), $m(i, j)$ the number of full sites between 
them, $\mathcal{N}(i, j)$ is the set of nearest-neighboring sites of $i$ 
which are neither $j$ nor a nearest-neighbor of $j$ --- $\mathcal{N}$ 
for \emph{not-maintained} ---, $N(i, j)$ their number and $n(i, j)$ the 
number of full sites between them, Fig.~\ref {LatRule}.

 For simplicity, we study only some of the processes where the rate of 
any jump, say from site $i$ to site $j$, depends only on $V(i)$, 
$m(i, j)$, $n(i, j)$, $M(i, j)$ and $N(i, j)$, but our approach can be 
extended. We consider the simplest situation, where there is no other 
interaction than exclusion (hard-core repulsion), to be the case where 
each agent has the same probability to attempt a jump in a given time 
interval, irrespective of the actual occupancy state of the surrounding sites; 
as a consequence, the rate of jumping of an agent is proportional to the inverse of the
number of possible jumps it can do (number of empty surrounding sites). Moreover, all 
possible jumps in the lattice have the same probability to occur, and at large 
scales, the occupation probability obeys a linear diffusion equation while 
the mean quadratic distance of each agent to its departure point 
grows linearly with time, as if agents would not interact at all and do 
a simple random walk~\cite{liggett-livre-1985, liggett-livre-1999}. In 
order to facilitate the comparison with results obtained on different 
lattices, we denote the rate of jump from site $i$ to site $j$ as
$T_{i, j}/V(i)$ with, by choice, $T_{i, j}=1$ when there is no other 
interaction than exclusion. The processes we consider (special forms of 
$T_{i, j}$ due to interactions) are listed below, along with their 
biological motivation.

\textbf{Adhesion model.} To study the influence of cell-cell adhesion on 
cellular migration, Khain \etal introduced a 
model~\cite{khain-et-al-cell-cell-adhesion-wound-healing} where 
\begin{equation}\label{Tadh} T_{i, j} = (1-q)^{m(i, j)+n(i, j)}, \end{equation} $q \in 
[0,1]$ being a constant parameter to quantify the strength of adhesion 
(from 0, no adhesion, like in standard Simple Symmetric Exclusion 
Processes (SSEP) to~1, impossible movement). It is assumed that adhesion 
is instantaneously gained or lost, \ie that the time scale of a possible 
dynamics of adhesion, like the recruitment of proteins to build up or 
strengthen focal adhesions, is much shorter than the time scale of 
migration.

\textbf{Gap junctional model.} 
 To explain experimental 
results about the influence of gap junction communications between cells 
on the migration of some tumoral 
astrocytes~\cite{aubert-et-al-migration-sans-astrocytes-sains, 
aubert-et-al-migration-avec-astrocytes-sains}, Aubert \etal introduced 
an exclusion process where \begin{equation} \label{Tgap}T_{i, j} = 1-p + (2p-1) \, 
\mathrm{min}[m(i, j), 1]. \end{equation} The parameter $p \in [0,1]$, assumed constant and common to all cells, allows to interpolate between 
SSEP for $p=1/2$, maximal effect of gap junctions for $p=1$, where no cell 
will move unless it has a neighbor cell and will keep contact with 
it, and $p=0$, where, to the contrary, no cell will move if it is not able 
to break all existing contacts (this case is probably of little 
relevance to biology, but was studied in detail in the context of the 
glassy dynamics~\cite{jaeckle-kroenig-2tlg-autodiffusion, 
kroenig-jaeckle-2tlg-diffusion-collective, pan-garrahan-chandler-2tlg, 
hedges-garrahan-2tlg}). Gap junctions are short channels passing through 
two touching cell membranes, which enable the passage of small molecules 
and ions. They are formed by two connected hemi-channels, one on each 
cell membrane. Different types of gap junctions arise from various 
genic expressions, and a gap junction can be functional or closed. The 
parameter $p$ should be a way to take into account such variability. 
Here again, it is assumed that the time scale of establishing effective 
gap junctions is much shorter than the time scale of migration and, 
additionally, that gap junctions are formed and open with any nearest 
neighbor cells and that their number has no influence (\ie maintaining 
communication with one cell has the same effect as maintaining it with 
many).

\textbf{Linear model.} To gain an overview, we finally introduce a more 
general model, where 
\begin{equation} \label{Tlin} T_{i, j} = \alpha + \beta \, m(i, j) 
+ \gamma \, n(i, j), 
\end{equation} 
$\alpha \ge 0$, $\beta$, and $\gamma \le 0$ being 
constant real numbers, Fig.~\ref {LatRule}. We choose a linear expression for $T_{i, j}$ to keep 
computations simple: the purpose of this model is not to be faithful to 
experiments, but to be illustrative. Choosing $\gamma=0$ and $\beta > 
0$, one gets a behavior similar to the gap-junctional model (the jump 
rate being now dependent on the number of cells with which contact is 
maintained). Choosing $\beta=\gamma < 0$, one gets a behavior similar 
to the adhesion model (for small $q$, $\alpha=1$ and $\beta=\gamma=-q$, 
the behavior should be quantitatively the same as for the adhesion 
model).

\subsection{Choice of boundary conditions}\label{bound}

The models previously proposed can be studied in different geometries, 
that is different boundary conditions and different numbers and disposition of cell sources and cell 
sinks, to address various aspects of the exclusion processes such as: 
relaxation to the equilibrium, steady state analysis, approach to the 
steady state, or perennial out-of-equilibrium conditions. One of the 
geometrical set-ups we will focus on is directly inspired by the 
set-up of the cancer cell migration process experiment discussed in 
\cite{limite-continue-modele-ma-champ-moyen, 
aubert-et-al-migration-sans-astrocytes-sains, 
aubert-et-al-migration-avec-astrocytes-sains}. The experiments consist 
in placing an aggregate of cancer cells (a so-called spheroid) in a 
Petri dish with an agar substrate containing suitable nutritional needs 
for the sustainment of the cells. Initially piled in the spheroid, the cells
slowly exit it and start to migrate in the outside region where they 
avoid overlapping. In the same way, we study the evolution of the system 
starting from a completely empty initial condition except for a small 
central region where all the cells are placed. In the proposed 
geometries, the spheroid is represented as a source of cells which can 
never empty out and where no empty sites are allowed at any time. When a 
cell leaves the source to enter the system, the free tile in the source, 
previously occupied by the cell, is immediately filled up with a new 
cell.

To avoid dealing with infinite lattices, 
we decided to add a sink region. It is an empty reservoir where cells 
are taken away from the system and act as if they were driven 
by a strong apoptotic signal putting them to death with an infinite 
apoptosis rate. Therefore, any cell entering the sink is destroyed. When 
the sink or the source extend over a set of contiguous sites that create 
a closed path, they become borders separating the space into independent 
subregions. In this work, for simplicity, we consider geometries where 
there is only one region of interest, always enclosed between a  source and 
a sink.

Although cells interact only with their nearest neighbors and travel 
during each elementary jump a length $\lambda$ equal to the distance between two 
neighbor cells, the boundaries 
may have influence on a long distance because of the exclusion rule. Actually, starting from an 
initially empty lattice, except for the source, there will be different 
time regimes: first a non-stationary period of time, when the population 
of cells invades the free space and the sink has not yet been reached, 
then an out-of-equilibrium steady state with a constant current of cells 
from the source to the sink (up to stochastic fluctuations). The first 
period of time is relevant for the migration experiments on Petri 
dishes.

\subfigcapmargin = 14pt
\begin{figure}[h]
\begin{center}
\subfigure[\hspace{2pt}]{\label{GeoCyl}
\includegraphics[width=0.467\textwidth, height=0.32\textwidth, angle=0]{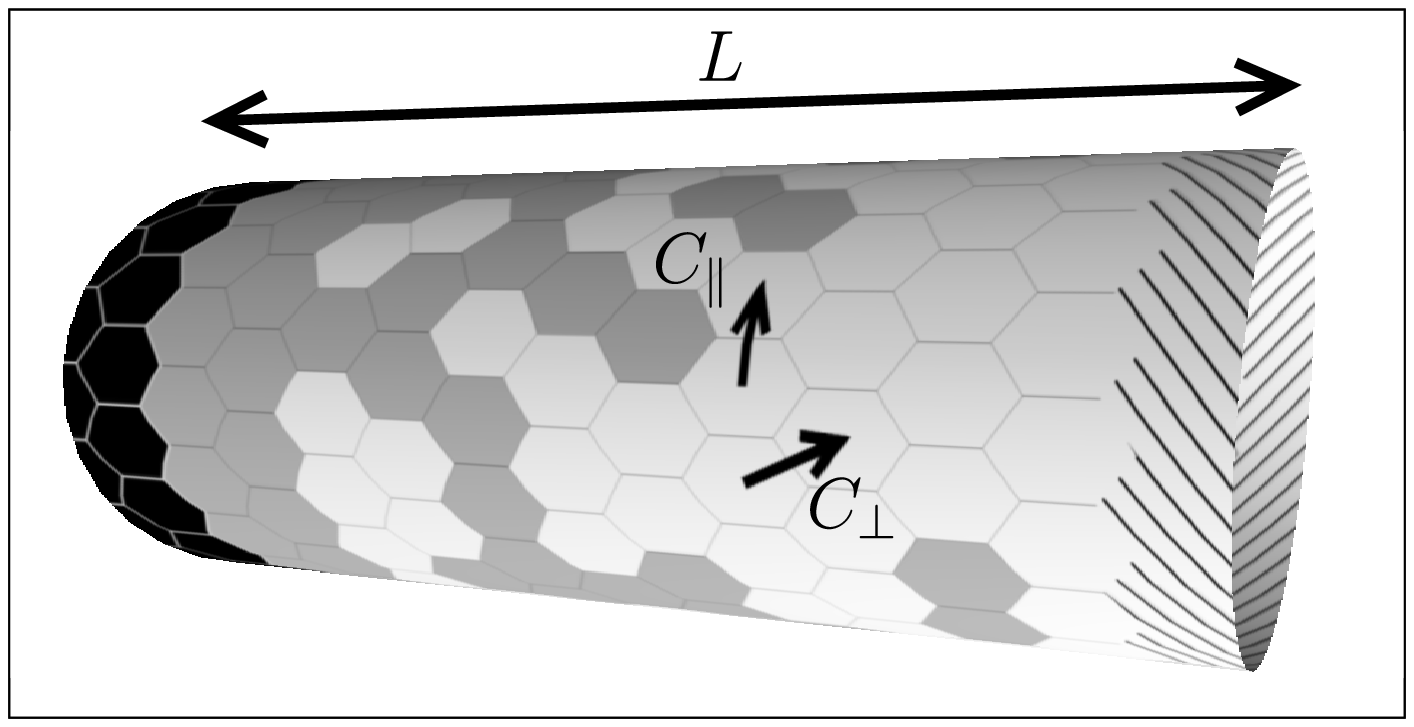}}
\subfigure[\hspace{2pt}]{\label{GeoRad}
\includegraphics[width=0.467\textwidth, angle=0]{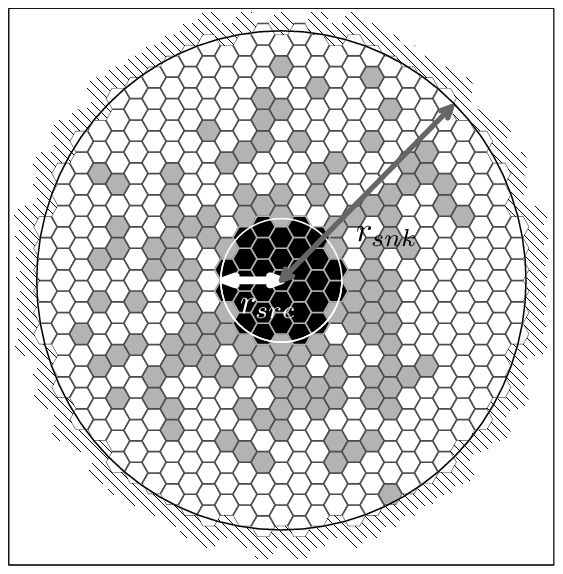}} 
\caption{Geometrical disposition of the reservoirs. The black sites represent the cell source, the hatched sites are the empty reservoir, the empty sites are white and the cells are in gray.  \subref{GeoCyl} cylindrical geometry of length L with two small arrows connecting neighbor sites showing the two main directions  of the correlation.   \subref{GeoRad} radial geometry with two concentric circumferences  showing the borders of the source and the sink with radii $r_{src}$ in white and $r_{snk}$ in gray respectively.\label{geoF}}
\end{center}
\end{figure}
\subfigcapmargin = 0pt

Given the set of rules specified in the previous section, we analyze  the dynamic evolution toward the steady state in two types of geometrical configurations. One on a cylindrical surface which has more theoretical advantages, and one on a planar bounded surface to mimic more closely in-vitro experiments of cells migration in culture dishes, see Fig.~\ref {geoF}.  
In the cylindrical configuration, the geometry consists of a two dimensional rectangular space with regular hexagonal tiling having periodic boundary condition 
along one direction and the other two sides in contact with two reservoirs: a source and a sink respectively, Fig.~\ref {GeoCyl}.
This geometry is particularly convenient. On one hand, it is invariant under translations along the direction with periodic boundary conditions. Thus, it makes it possible to describe the  properties of the two dimensional system  such as density and correlations in terms of just one single spatial variable: for example, the distance from one reservoir. On the other hand, this ``quasi-1D'' geometry does not generate a loss of generality. More complex geometries can be equally addressed with the techniques of this work, and they do not introduce further physical effects. At least at large times, the results are fundamentally the same up to some  numerical changes (see Sec.~(\ref{sec:level3D})  for results on a different geometry and Sec.~(\ref{sec:level4c})  for a justification). In the cylindrical geometry, the shape of the reservoirs are fixed and the degrees of freedom are the two sizes of the cylindrical surface: the length of the circumference and the distance between the reservoirs. From numerical simulations, we have seen that the results are independent of the length of cylindrical circumference, if this is 10 or more lattice tiles $\lambda$ (results not shown). The distance between the reservoirs produces some differences in the steady state, and they will be discussed in the next section. It is important to remark that these considerations hold true when the lattice is oriented as in  Fig.~\ref {GeoCyl} as well as when it is rotated by 90 degrees.   
The second geometrical configuration is a planar 
regular hexagonal tiled lattice, hereafter called radial geometry. In it, we define an origin $O$ represented by a generic tile and two radii from $O$: $r_{src}$ and $r_{snk}$. The tiles with distance from $O$ less than or equal to $r_{src}$  function as source and the tiles with distance from $O$ bigger than $r_{snk}$ function as sink, Fig.~\ref {GeoRad}. The only constraint that exists between the two degrees of freedom,  $r_{src}$ and $r_{snk}$, is 
$r_{src}< r_{snk}$. Nevertheless, numerical results prove that for radii larger than two lattice steps $\lambda$, their specific values become weakly influential on the dynamical evolution of the system (results not shown). Also, in this geometrical configuration, the steady state  can depend on the distance between the reservoirs.

\section{\label{sec:level3} Evolution equations on the lattice}

 In \cite{effet-inhibition-jonctions-communicantes}, the authors analyzed similar systems in the mean-field approximation, and in \cite{limite-continue-modele-ma-champ-moyen}, they commented about the discrepancy at the steady state between the numerical and the analytical results for 
particular values of their interaction parameter close to the sink. 
On the other hand, here, we address the problem beyond the mean-field approximation to investigate the behaviors of the correlation at short distance among cells defined as: 
\begin{equation}\label{C2}
C_2(i, j)=\langle\eta(i)\eta(j)\rangle-\langle\eta(i)\rangle\langle\eta(j)\rangle,
\end{equation}
where $i$ and $j$ are two nearest neighbor sites, $\eta$ is the number of cells, and the angular brackets stand for the average over all possible configurations. 
 
\begin{figure}
\includegraphics[width=0.34\textwidth, angle=-90]{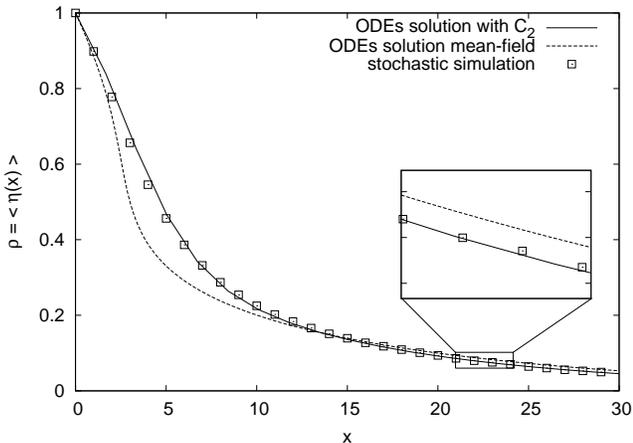}
\caption{\label{comparison} Density profile for the adhesion model with $q=0.4$ on a cylindrical geometry after 80 time steps. The dots represent the results of the stochastic simulation of Sec.~(\ref{sec:level3D}), and the error bars are smaller than the size of the symbols. The continuous line shows the density profile obtained by considering the two-point nearest neighbor correlation functions as explained in Sec.~(\ref{subsec:level3b})  and Sec.~(\ref{subsec:level3c}). The dashed line is the density profile in the mean-field approximation.}
\end{figure}

In addition, considering the presence of correlations between nearest neighbors results in a more accurate prediction for the density profile in comparison to that obtained in the mean-field approximation. For example, this is evident in the adhesion model when the interactions are sufficiently strong, as shown in Fig.~\ref{comparison}.
 
It is important to stress that the previously stated rules and the results  shown in the next sections will hold true not just as a consequence of the particular choice of the hexagonal tiling. Indeed, the phenomenon of a correlation wave with similar behaviors is a much more general result, and it will be present in other kinds of lattices, for example: triangular, square, and  random tilings, under the condition of a more general definition of nearest neighbors like, for example,  ``a tile is a nearest neighbor of another if they have a common vertex''. 

\subsection{Master equation and evolution equations for the correlation 
functions --- the general case}
Let us define a lattice as a partition of a $n$-dimensional space $\Omega\subseteq\mathbb{R}^n$  in $z$ non-overlapping subsets,
each of them representing a tile of the lattice identified by an index $i\in \mathbb{N}$. 
Let $\eta_i$ be the number of cells in lattice site $i$; $\eta_i \in 
\{0,1\}$ so that $\eta_i^2=\eta_i$, and the total number of cells in the 
system at time $t$ is: $Z=\sum_{i=1}^z\eta_i$. 
The generic configuration of the occupancy states is given by
vector $\bs{\eta}=(\eta_1, \eta_2, \ldots, \eta_z)$ of all the 
numbers $\eta_i$. We denote $P(\bs{\eta}, t)$ the probability that the 
the process is in the configuration $\bs{\eta}$ at time $t$.

If $\hat{W}_{i, j}$ is the operator that permutes the contents of sites 
$i$ and $j$: $$\hat{W}_{i, j}: 
(\eta_1,...,\eta_i...,\eta_j,...,\eta_z)\rightarrow 
(\eta_1,...,\eta_j,...,\eta_i,...,\eta_z),$$ we can express the 
evolution equation for $P(\bs{\eta}, t)$ (the master equation) as
\begin{align}\label{ME}
\partial_t 
P(\bs{\eta}, t) & =  \sum_{i=1}^z \frac{1}{V(i)} \mathop{\sum_{j\in\mathcal{V}(i)}} 
\left[ 
 \eta_j(1-\eta_i) T_{j, i} \left( \hat{W}_{i, j} \bs{\eta} \right)
 \times
 \right. \nonumber \\ 
 & P \left( \hat{W}_{i, j} \bs{\eta}, t \right) 
 \left. - \eta_i (1-\eta_j) T_{i, j}(\bs{\eta}) P(\bs{\eta}, t) \right] 
\end{align} 
where the first term inside the brackets represents the 
flux of probability from the configurations which contribute to fill the 
site $i$, so that $\eta_j(1-\eta_i) T_{j, i} \left( \hat{W}_{i, j} \bs{\eta} 
\right) \mathrm{d} t$ is the chance of a cell to move from $j$ to $i$ in 
the infinitesimal time step $\mathrm{d} t$ provided that the system is 
in the state $\hat{W}_{i, j} \bs{\eta}$. Similarly, the second term 
represents the flux of probability to the configurations where the site 
$i$ is empty. This expression is valid for any exclusion process; we 
assume from now on that $T_{i, j}(\bs{\eta})$ vanishes if $i$ and $j$ are 
not nearest neighbors on the lattice (\ie if $j \notin 
\mathcal{V}(i)$).

 We define the average of generic quantity $A(\bs{\eta})$ at time $t$ as a 
sum over all possible configurations of the process: \begin{equation} 
\langle A \rangle = \sum_{\bs{\eta}} A(\bs{\eta}) P(\bs{\eta}, t). 
\end{equation} In particular, we are interested in the average 
multi-point density, or correlation function, on $n$ distinct lattice 
sites $l_1, l_2, \ldots, l_n$: 
\begin{equation}\label{multiMean} 
\rho_n(l_1,\ldots, l_n, t)= \langle \eta_{l_1} \eta_{l_2} \ldots 
\eta_{l_n} \rangle 
\end{equation} 
--- the density $\langle \eta_i 
\rangle$ of cells on a single lattice site $i$ will simply be written 
$\rho(i, t)$ --- and in the connected correlation function: 
\begin{align}\label{defconncorrfun} C_n(l_1,\ldots, l_n, t) & =  
\left\langle [\eta_{l_1} - \rho(l_1, t)] [\eta_{l_2} - \rho(l_2, t)] 
\ldots \right. \nonumber \\ &  \left. \ [\eta_{l_n}-\rho(l_n, t)] 
\right\rangle \end{align} 
which vanishes if the occupation numbers on 
the sites are statistically independent.

 Inserting the master equation Eq.~(\ref {ME}) into the expression of the 
time derivative of $\rho(i, t)$ yields the evolution equation:
\begin{align}\label{rho}
\partial_t\rho(i, t)&=\frac{1}{V(i)}\left\langle\mathop{\sum_{\phantom{,}j\in\mathcal{V}(i)}}[T_{j, i}(\bs{\eta})\eta_j(1\!-\!\eta_i)+\right.\\
 & \left. \phantom{\sum_{k}}  -T_{i, j}(\bs{\eta})\eta_i(1-\eta_j)]\right\rangle.
\end{align}
Using the general definition in Eq.~(\ref {multiMean}) for $n=2$, the equation for the time evolution of the two-point density function at the two generic sites $i$ and $j$ is \cite{erban}: 
\begin{align}\label{rho2}
\partial_t\rho_2(i, j,t)=\frac{1}{V(i)}\left\langle  
\mathop{\sum_{k\in\mathcal{V}(i)}}_{k\neq j}
[T_{k, i}(\bs{\eta})\eta_k(1-\eta_i)
+\right.\nonumber\\
\left.\phantom{\mathop{\sum_{k\in\mathcal{V}(i)}}_{\textrm{and}\,k\neq j}}
-T_{i, k}(\bs{\eta})\eta_i(1-\eta_k)]\eta_j\right\rangle
+\,i\leftrightarrow j,\end{align}
where $i\leftrightarrow j$ is equal to the first term on the right-hand side of the equation above, with the roles of $i$ and $j$ exchanged.
In Eq.~(\ref {rho2}), the constraints $k\neq j$ are added to ensure that all 
configurations included in the counting after a jump of a cell have both sites $i$ and $j$ occupied.   
The equations for the two point connected correlation function 
immediately follow from~Eqs.~(\ref {defconncorrfun}, \ref {rho}, \ref {rho2}):
\begin{align}\label{C2def}
& \partial_tC_2(i, j,t)=\frac{1}{V(i)}\left\langle  \mathop{\sum_{k\in\mathcal{V}(i)}}_{k\neq j}
[T_{k, i}(\bs{\eta})\eta_k(1-\eta_i)+\right.\nonumber\\
& \left.\phantom{\sum_{k}}
-T_{i, k}(\bs{\eta})\eta_i(1-\eta_k)](\eta_j-\rho(j, t))
\right\rangle +\,i\leftrightarrow j.	
\end{align}  

In the same way, we can express the evolution equation of connected 
correlation functions for any $n$.

\subsection{\label{subsec:level3b}  Expressions for our models}

 Let us give the expression of $T_{i, j}(\bs{\eta})$ for the models we 
introduced in Sec.~(\ref {sec:level2}), on the hexagonal tiling.
To explicitly compute the averaged quantities of the previous subsection, we need to express 
$m(i, j)$ and $n(i, j)$ in terms of $\bs{\eta}$: 
\begin{equation} m(i, j) = \sum_{k \in \mathcal{M}(i, j)} \eta_k, \qquad 
n(i, j) = \sum_{k \in \mathcal{N}(i, j)} \eta_k. \end{equation}

Consequently, the transition rate for the adhesion model in Eq.~(\ref {Tadh}) becomes: $$T_{i, j}(\bs{\eta}) = 
(1-q)^{\eta_k + \eta_l + \eta_r + \eta_s + \eta_w},$$ 
where $k$ and $l$ are the common neighbors 
of $i$ and $j$, and $r, s$ and $w$ are the neighbors of $i$ which are not 
neighbors of the future position $j$.
In the gap junctional model, when intercellular communication 
through gap junctions drives the system dynamics, Eq.~(\ref {Tgap}) can be rewritten as:  
$$T_{i, j}(\bs{\eta})=p(\eta_k+\eta_l-\eta_k\eta_l)+(1-p)(1-\eta_k)(1-\eta_l),$$ 
where the sites $l$ and 
$k$ identify the two common neighbors of $i$ and $j$.  In this way, 
$T_{i, j}$ will be equal to $p$ if both, or just one of the sites $k$ 
and $l$ are occupied by other cells, and it will be equal to $1-p$, if 
both $k$ and $l$ are empty.  When $\eta_i=1$, the cell in $i$ will share 
gap junctional links with all the nearest neighbor sites occupied by 
other cells, but only the gap junctions in the direction of the site 
$k$ or $l$ will be maintained functional during the cell transition to 
the site $j$.
Finally, in the linear model  the transition rate Eq.~(\ref {Tlin}) is: 
$$T_{i, j}(\bs{\eta})=\alpha+\beta(\eta_k+\eta_l) + \gamma 
(\eta_r + \eta_s + \eta_w),$$ 
where the indices $k, l, r, s$ and $w$ have the same meaning as in the adhesion model.

For the gap junctional model and for the linear model with $\gamma=0$, the 
transition rate is invariant under the permutation of indices $i$ and 
$j$: $T_{i, j}(\bs{\eta})=T_{j, i}(\bs{\eta})$. For the 
adhesion model, it is not invariant because the sites belonging to $\mathcal{N}(i, j)$ in the 
expression for $T_{i, j}(\bs{\eta})$ are different from the  sites belonging to $\mathcal{N}(j, i)$  
in $T_{j, i}(\bs{\eta})$. When $T_{i, j}(\bs{\eta})$ is symmetric with respect to the indices $i$ and $j$, the probability distribution 
of the configurations at equilibrium is the uniform distribution, so 
that, at equilibrium, the occupation of one site is statistically 
independent from the others.

In the presence of sources and sinks, the general evolution equations for the densities and the correlation given in Eqs.~(\ref {rho}, \ref {rho2}, \ref {C2def}) no longer hold. Indeed, if one or more points of the multi(single)-point density functions are at one lattice step from, or belong to a reservoir, it is necessary to take into consideration that some changes in the configurations of the positions of cells from, or toward the reservoirs are impossible and must be excluded.
For example, when a multi(single)-point density function has a point at distance $\lambda$ from a sink, 
no flux of cells coming from the sink contributes to the final configuration. On the other hand, if the reservoir is a source, no cell can enter it and  any flux of cells toward the source must be zero.   % 

From Eq.~(\ref {defconncorrfun}), it follows that if one of the $\eta_i$ is constant, then the correlations including the site $i$ are zero; consequently, it is easy to express the border conditions in terms of the multi-point correlation functions with one or more of its points belonging to a reservoir. It is important to stress that, despite the presence of reservoirs that invalidates the general forms of Eqs.~(\ref {rho}, \ref {rho2}, \ref {C2def}),  substituting the values of the border condition into the evolution equations of the density and the correlations explicitly derived from Eqs.~(\ref {rho}, \ref {C2def}) produces the same correct results as if they were obtained from the evolution equations with the ulterior constraints due to the presence of reservoirs. 

Let us consider the case of the linear model and explicitly compute the equations for the density and the two-point connected function in the hexagonal lattice.
Using Eqs.~(\ref {rho}, \ref {C2def}), and dropping the  explicit dependence of the densities on time and unnecessary indices,  the system of equations, in the region of interest far from the reservoirs, is:

\begin{widetext}
\begin{align}\label{sys}
\partial_t&\rho(i)=\frac{1}{6}\mathop{\sum_{k\in\mathcal{V}(i)}}\left\{ 
\alpha [\rho(k)-\rho(i)]+\beta 
\sum_{s\in\mathcal{M}(i, k)}[C_2(k, s)+\rho(k) 
\rho(s)-C_2(i, s)-\rho(i)\rho(s)]+\right.\nonumber\\
& \left.\qquad\qquad+\gamma \sum_{s\in\mathcal{N}(i, k)}[\rho(i, k,s) 
-\rho(i, s)]-\gamma\sum_{s\in\mathcal{N}(k, i)}[\rho(k, i,s) 
-\rho(k, s)] \right\}, \nonumber\\
&\partial_t C_2(i, j)=\frac{1}{6}\mathop{\sum_{k\in\mathcal{V}(i)}}_{k\neq 
j}\left\{ \alpha \Big[C_2(k, j)-C_2(i, j)\Big]+ 
\beta\sum_{s\in\mathcal{M}(i, k)}
\bigg[\rho(j, k,s)-\rho(j)\rho(k, s)-\rho(j, i,s)+\rho(j)\rho(i, s)\bigg]+
\right.\nonumber\\&  
\qquad \quad +\gamma \sum_{s\in\mathcal{N}(i, k)} 
\bigg[\rho(j, i,k, s)-\rho(j)\rho(i, k,s)-\rho(j, i,s)+\rho(j)\rho(i, s)\bigg] +
\nonumber\\ &  \left. \qquad \qquad
-\gamma\sum_{s\in\mathcal{N}(k, i)}\bigg[\rho(j, k,i, s)-\rho(j)\rho(k, i,s)
-\rho(j, k,s)+\rho(j)\rho(k, s)\bigg] \right\} +i\leftrightarrow j,
\end{align}
\end{widetext}
\begin{widetext}
\begin{align}
\rho&(i, j,k)-\rho(i)\rho(j, k)=C_2(i, j)\rho(k)+C_2(i, k)\rho(j)+C_3(i, j,k),\label{pippe}\\
\nonumber\\
\rho&(i, j,k, l)-\rho(i)\rho(j, k,l) =  C_2(i, j)\rho(k)\rho(l) + 
C_2(i, k)\rho(j)\rho(l) + C_2(i, l)\rho(j)\rho(k) +
\nonumber \\ & \qquad\qquad
+C_3(i, j,k)\rho(l) + C_3(i, j,l)\rho(k) + C_3(i, k,l)\rho(j) + C_4(i, j,k, l),\label{pippe2}
\end{align}  
\end{widetext}
where the Eqs.~(\ref {pippe}, \ref {pippe2})  show how to rewrite the quantity  $\rho(i, j,k)$ and $\rho(i, j,k, l)$  for the generic sites $i,j,k$ and $l$ in terms of connected correlation functions.  

The border conditions for the system Eq.~(\ref {sys}), at any time $t$, are:
\begin{equation}
\begin{array}{ll}
\rho(l_i)=1,   &\;\;\textrm{ for } l_i \textrm{ in the source}\\
\rho(l_i)=0,   &\;\;\textrm{ for } l_i \textrm{ in the sink}\\
C_n(l_1,\ldots, l_n)=0, & \;\;\;\textrm{if any }\, l_i  \textrm{ is in a reservoir}.
\end{array}
\label{BC}
\end{equation}
The same set of border conditions Eq.~(\ref {BC}) can be applied in both cylindrical and radial geometrical dispositions of source and sink, and in addition, they hold true for all three models proposed.

For a lattice with $z$ tiles, in the system of Eqs.~(\ref {rho}, \ref {C2def}), there are $z$ differential equations for the density  and even more for the two-point connected correlation function. If we consider only the correlations between nearest neighbor sites, the process is described by a total of $(1+d)z$ ordinary differential equations plus the border conditions, where $d \times z$ equations describe the time evolution of the two-point connected correlations in the $d$ main lattice directions (for the hexagonal tiling $d=3$). 
In the cylindrical geometry, due to the invariance under rotations and reflections along the axial direction, for a site there are only two independent equations for the correlations between nearest neighbor sites: $C_\parallel$ and $C_\perp$, see Fig.~\ref {GeoCyl}; therefore, the number of equations reduces to $4 L$, where $L$ is the minimum number of sites one has to travel through to go from one end to the other of the cylinder.

\subsection{\label{subsec:level3c} Closure approximation}

To solve the set of equations Eqs.~(\ref {sys}, \ref {BC}) for the linear 
model, it is necessary to express $C_3(i, j,k, t)$ in terms of known 
quantities. In more general cases such as the gap junctional and the 
adhesion models, finding a solution of the system of equations 
Eqs.~(\ref {rho}, \ref {C2def}, \ref {BC}) requires to already know all the 
correlations $C_n(l_1,\ldots, l_n)$ with $n>2$, or to add other equations 
to the initial system which allows one to determine these unknown 
quantities. One possible way is to express the multi-point correlation 
functions with the highest number of points with an approximate 
expression involving only multi-point correlation functions with less 
points. This approach is called closure approximation. In comparison to 
the approach in articles \cite{limite-continue-modele-ma-champ-moyen, 
effet-inhibition-jonctions-communicantes}, where mean-field 
approximation was adopted and all the correlations were completely 
disregarded, we take into consideration the short range two-point 
connected correlation functions with the purpose to obtain more 
information about the dynamical evolution of the system  and to improve the 
agreement of the analytical results with the stochastic simulations. Our 
closure approximations is the following. All the connected correlations 
defined on more than two points are set to zero: 
\begin{equation}\label{CApp1} \left\langle\prod_{k=1}^{n>2} 
(\eta_{l_k}-\langle\eta_{l_k}\rangle)\right\rangle=0, \end{equation} 
where, as above, $l_1, l_2, \ldots, l_n$ are distinct lattice sites, and 
all the two-point connected correlation functions between cells at a 
distance bigger than one lattice step $\lambda$ are set to zero: 
\begin{equation}\label{CApp2} \left\langle \left(\eta_i - \langle \eta_i 
\rangle \right) \left( \eta_j - \langle \eta_j \rangle \right) 
\right\rangle = 0, \qquad \textrm{if} \ j \notin \mathcal{V}(i), j \neq i. 
\end{equation} 
Therefore, only the information relative to the nearest 
cell couples remains, as if all the clusters and structures with more 
than two points would appear in a completely random way (conditioned to 
the values of the local density of cells and of the two-point 
nearest neighbor correlation functions).

These approximations are suggested by the rules of our exclusion 
processes, in which the movement of one cell is directly influenced only 
by the presence or absence of nearest neighbors. As we shall see, 
stochastic simulations show that the approximations on the system 
Eqs.~(\ref {rho}, \ref {C2def}) produced by Eq.~(\ref {CApp1}) are reasonably good 
for the gap junctional model, the linear model when $\gamma$ is small, 
and the adhesion model when $q$ is small (weak adhesive interactions). 
For instance, simulations of the gap junctional model show that, except 
cell couples moving together, no particular structure or big cluster 
appears.
Nevertheless, neglecting correlations 
between sites at distances of two lattice steps and above is not so 
unquestionable because of both ``repulsive'' interactions (exclusion) and 
``attractive'' interactions (adhesion). On one hand, it is known that, 
in exclusion processes with no other interaction than mere exclusion, 
there are long range 
correlations~\cite{spohn-correlations-hors-equilibre, 
garrido-lebowitz-maes-spohn-correlations-hors-equilibre, 
spohn-livre-limite-hydrodynamique}. From stochastic simulations it seems 
that correlations between cells at two and three lattice steps apart are 
much smaller, but not completely negligible (results not shown). In fact, the 
time evolution of these correlations and their behaviors resemble those 
of the connected correlation functions between nearest neighbors. On 
the other hand, our models with large values of the adhesion parameter 
($q$ or $\gamma$) exhibit large-scale structures of cells. For instance, 
the adhesion model can be mapped onto the Ising 
model~\cite{khain-et-al-adhesion-lignees-de-gliomes-sans-migration}, and 
it can be shown that a spontaneous phase separation with clustering of 
cells happens at larges values of $q$. In such situations, the success 
of the analytical approach in reproducing exactly results from the 
stochastic simulations is lost. But, at least when aggregation is absent 
or weak, it is still possible to retrieve, from the analytical results 
for the connected correlation functions, important information regarding 
the system evolution which is characteristic of each type of interaction 
(see Sec.~(\ref {sec:level3D}) and Sec.~(\ref {sec:level4})).

The closure approximations, Eqs.~(\ref {CApp1}, \ref {CApp2}) can be 
shortly expressed together with the set of border conditions 
Eq.~(\ref {BC}):
\begin{equation}
\begin{array}{ll}
\rho(l_i)=1,   &\;\;\textrm{ for } l_i \textrm{ in the source}\\
\rho(l_i)=0,   &\;\;\textrm{ for } l_i \textrm{ in the sink}\\
C_n(l_1,\ldots, l_n)=0, & \;\;\;\forall\, l_i  \textrm{ in a reservoir or } \forall\; n>2\\ 
C_2(l_1, l_2)=0, & \;\;\;\forall\, l_i  \textrm{ in a reservoir or } l_1\notin \mathcal{V}(l_2).
\end{array}
\label{BC2}
\end{equation}

Other kinds of closure approximations can be applied in place of 
Eq.~(\ref {CApp1}). One can systematically extend this approximation by 
dealing with correlations at larger distances or with more than two 
sites~\cite{simpson-baker-correcting-mean-field-uniform-in-space, 
baker-simpson-correcting-mean-field-spatially-dependent, 
simpson-baker-correcting-mean-field-adhesion-1d}. One can also choose a 
different scheme, as the Kirkwood Superposition Approximation used in 
\cite{baker-simpson-correcting-mean-field-spatially-dependent, 
simpson-baker-correcting-mean-field-adhesion-1d}. However, this results 
in dealing with unbounded connected correlation functions which diverge 
when all sites are empty or when all are full, and do not satisfy the 
geometrical and initial conditions proposed here.

\subsection{\label{sec:level3D} Stochastic simulations and discrete equations }

The stochastic simulations consist of a cellular automaton 
where cells evolve through a number of discrete time steps moving  on a hexagonal tiling and performing a series of interaction dependent moves described by the rules in Sec.~(\ref {sec:level2}), \cite{asmussen2007stochastic}. 
In the framework of cellular automata, the rules and the change of the state of the cells are intended to be applied in parallel (synchronously) 
\cite{cellular-automata-complexity};
 nevertheless, 
in the proposed exclusion processes, the parallel update scheme entails  the problem of two cells jumping at the same time on the same empty site. To avoid any ambiguities in the stochastic simulations, the positions  of the cells  are asynchronously  updated following a random order time scheme \cite{random-order}: 
at each time step  all the cells are chosen once in a new random order  and updated. 
At equilibrium, for the gap junctional model and linear model with $\gamma=0$,  the symmetry of  $T_{i, j}=T_{j, i}$  results in no correlations, see Sec.~(\ref {subsec:level3b}). On the other hand, simulations with cyclic update scheme (all cells, at each time step, are updated once in the same random pre-fixed  order) produce spurious correlations at equilibrium. Therefore, the random ordered scheme (all cells, at each time step, are updated once in a new random order) \cite{ca-biomath} and the random independent scheme (at each time step, a cell is chosen randomly and updated) \cite{async-ca}\cite{ASEP-update} are more appropriate for the simulations.
Also, the clocked random waiting time scheme (each cell is updated following its own independent random clock) \cite{10-lectures-particle-sys} should not introduce spurious correlations,  
if the waiting time distribution has the first two moments finite. 

The  update of a generic cell in the site $i$ consists in choosing the new site  $j$ at random with equal probability between all the nearest neighbors $V(i)$, and  then, if the site $j$ is empty, moving the cell in the new position with a probability $Q= T_{i, j} (\bs{\eta}) \Delta t$. 
To be well defined, the probability $Q$ that a cell jumps during the time interval $\Delta t$ requires to fix the simulation time step consistently. Therefore, we choose:
\begin{equation}
\Delta t=
\begin{cases}
1 & \textrm{adhesion model}\\
1 & \textrm{gap junctional model}\\
\frac{1}{ \displaystyle{\max_{\bs{\eta^\prime}}}(T_{i, j}(\bs{\eta^\prime}))} \quad & \textrm{linear model}
\end{cases}
\end{equation}

where, in the linear model on the hexagonal tiling,   the normalization factor $ \max_{\bs{\eta^\prime}}
(T_{i, j}(\bs{\eta^\prime})) =\alpha+\max (2\beta,0)$ and the parameters must satisfy the inequality  $\alpha\geq -3\gamma-\min(2\beta,0)$.

To reduce the stochastic noise, 
we averaged the outcomes of a series of independent stochastic simulations with the same initial conditions for each model and each parameter relative to the interaction and geometry. 
The system of ODEs, Eqs.~(\ref {rho}, \ref {C2def}, \ref {BC2})  with the initial conditions corresponding to the CA simulations:
\begin{equation}\label{IC}
\begin{array}{ll}
\rho(l_i, t=0)=0,   &\;\;\textrm{ for } l_i \textrm{ not in the source}\\
C_2(l_1, l_2, t=0)=0, & \;\;\;\forall\, l_i,
\end{array}
\end{equation}
are numerically integrated using a fourth order Runge-Kutta method and compared with the results of the CA.

Let us show  and discuss in detail the results for the more interesting cases.

\begin{figure}[b]
\begin{center}
\subfigure[\hspace{2pt}]{
\includegraphics[width=0.35\textwidth, angle=-90]{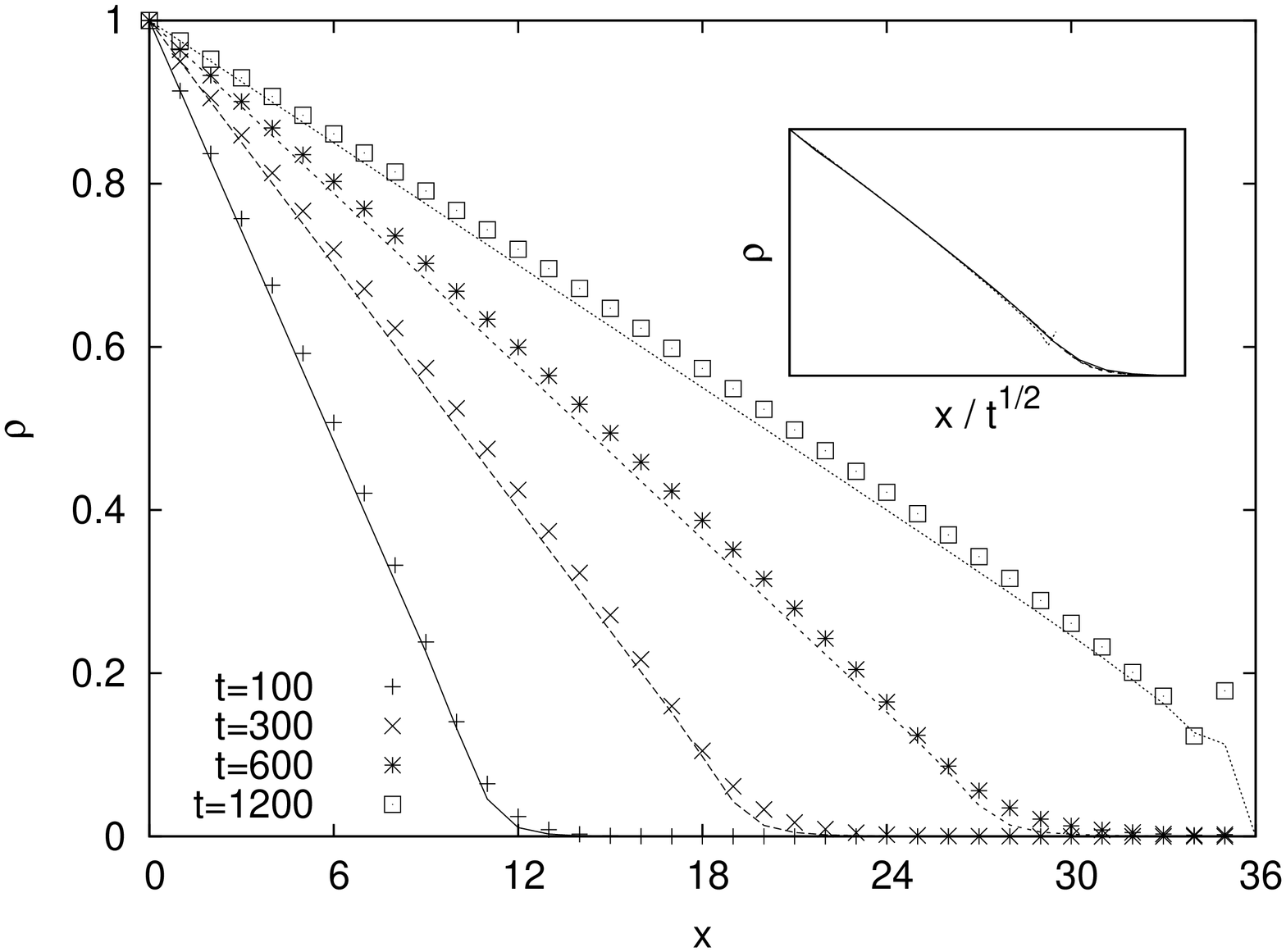}\label{GapDenCyl1}}
\vspace{6pt}

\subfigure[\hspace{2pt}]{
\includegraphics[width=0.35\textwidth, angle=-90]{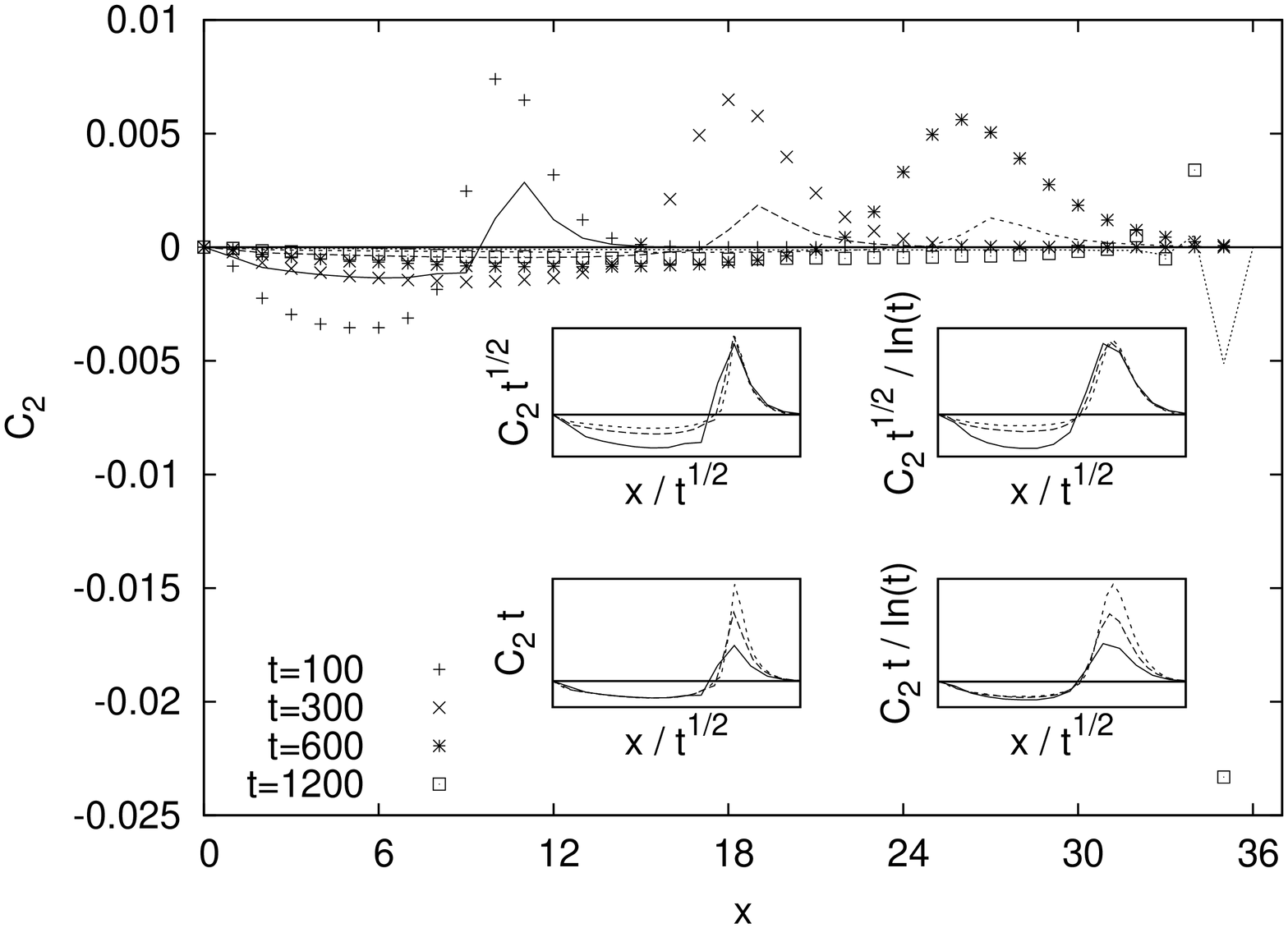}\label{GapCorCyl1}}
\vspace{6pt}
\caption{Profile of \Subref{GapDenCyl1} the density and \Subref{GapCorCyl1} the connected correlation at different times for the gap junctional model on a cylindrical geometry with p=1 and L=36. The dots represent the results of the stochastic simulations, and the error bars are smaller than the size of the symbols. The lines refer to the results of Eqs.~(\ref {rho}, \ref {C2def}, \ref {BC2}). The insets show the scaling behaviors. In \Subref{GapCorCyl1}, the insets on the left-hand side refer to the analytical results, and those on the right are the respective results obtained from cellular automata simulations. The scaling behaviors of the peak and on the negative tail are shown on the top row  and  bottom row of insets respectively.}\label{GapCyl1}
\end{center}
\end{figure}

\begin{figure}
\begin{center}
\subfigure[\hspace{2pt} ]{
\includegraphics[width=0.35\textwidth, angle=-90]{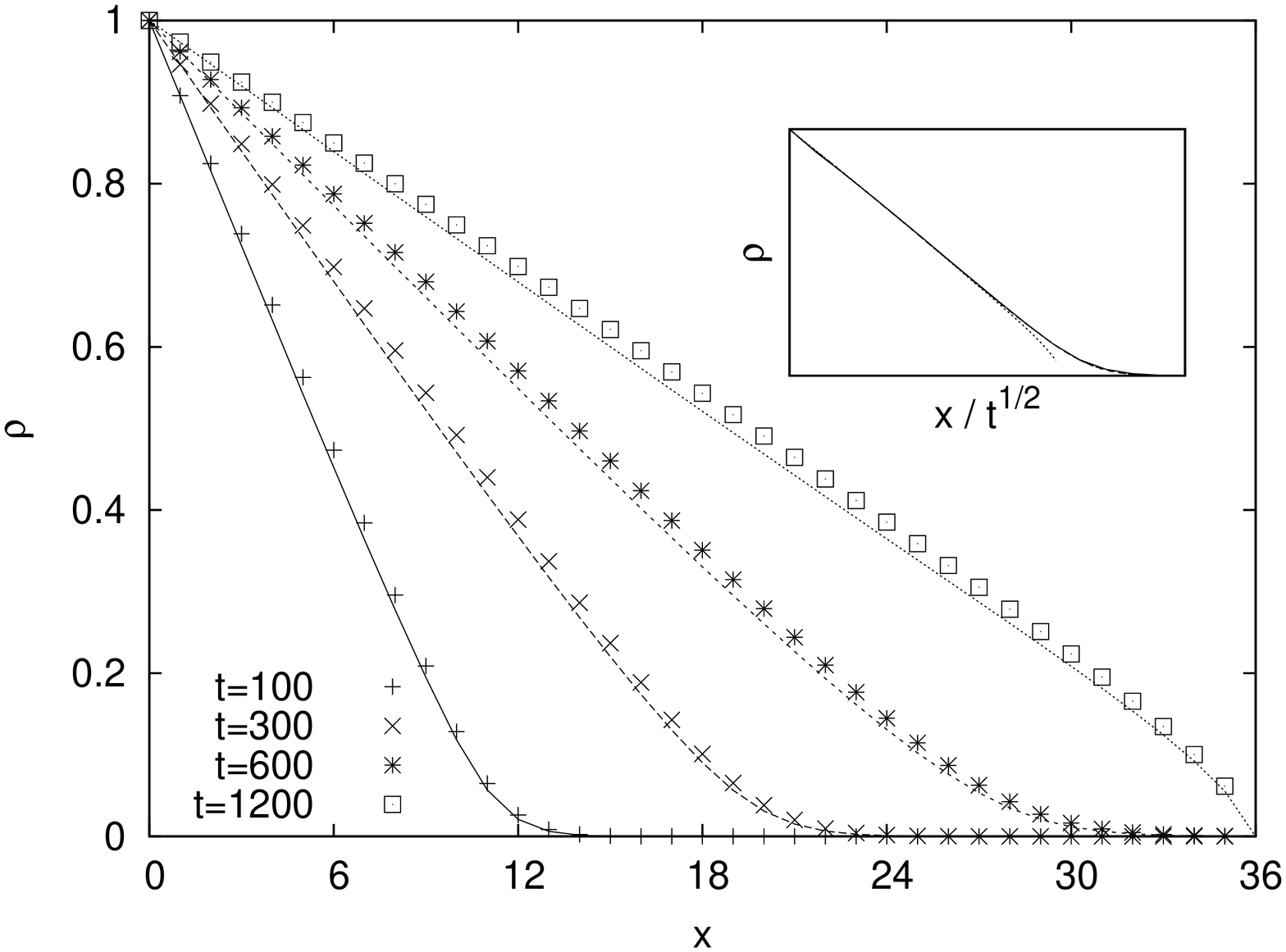}\label{GapDenCyl9} }
\vspace{6pt}

\subfigure[\hspace{2pt} ]{
\includegraphics[width=0.35\textwidth, angle=-90]{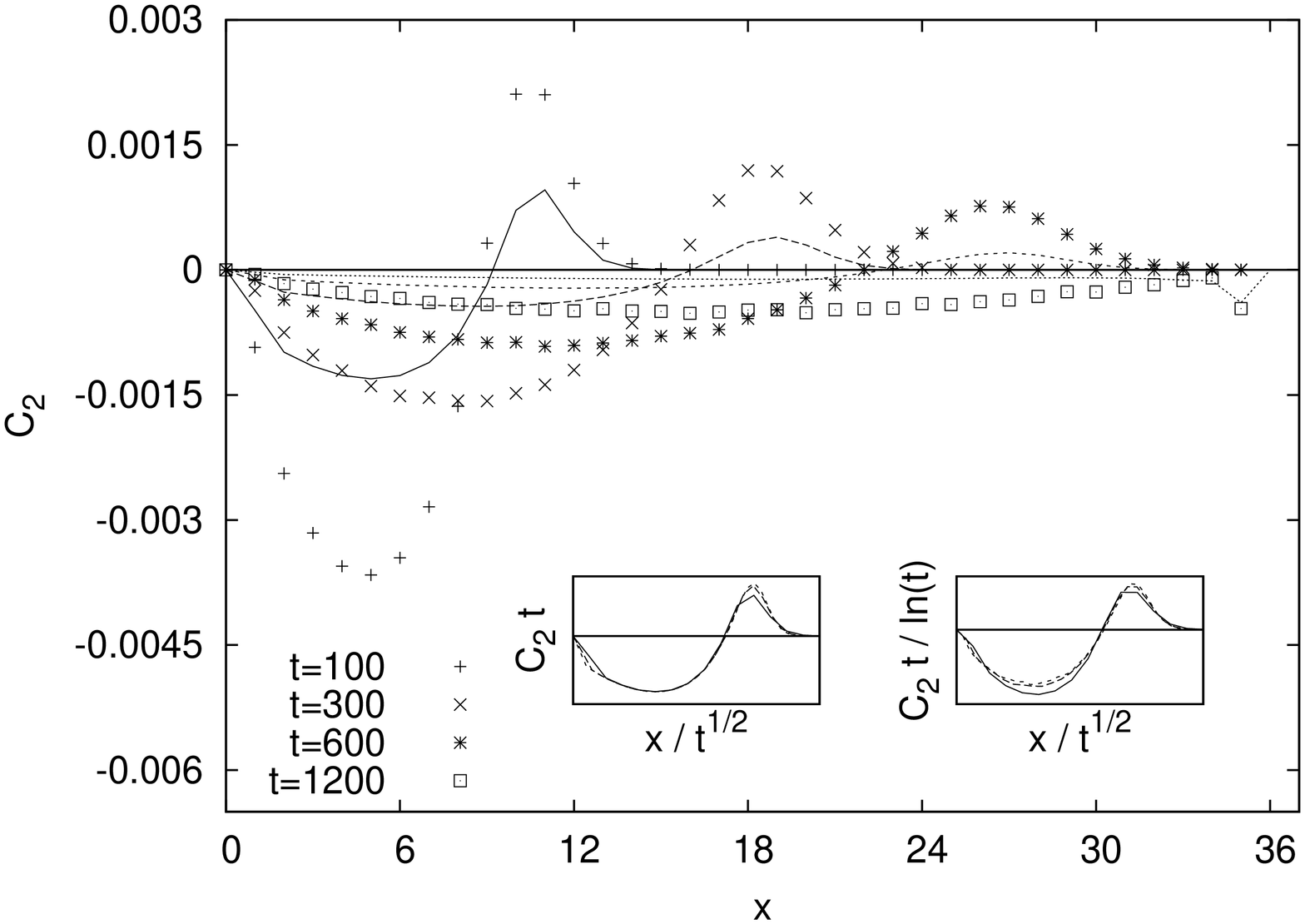}\label{GapCorCyl9} }
\vspace{6pt}
\caption{ 
Profile of \Subref{GapDenCyl9} the density and \Subref{GapCorCyl9} the connected correlation at different times for the gap junctional model on a cylindrical geometry with p=0.9 and L=36. The dots represent the results of the stochastic simulations, and the error bars are smaller than the size of the symbols. The lines refer to the results of Eqs.~(\ref {rho}, \ref {C2def}, \ref {BC2}). The insets show the scaling behaviors. In \Subref{GapCorCyl9}, the inset on the left-hand side refers to the analytical results, and that on the right shows the respective results obtained from cellular automata simulations.}\label{GapCyl9}
\end{center}
\end{figure}

\textbf{Gap junctional model --- density profile.}
The density profile at different times on a cylindrical geometry for the gap junctional model  with interaction parameter $p=1$ and $p=0.9$ are shown in Figs.~\ref {GapDenCyl1}, \ref {GapDenCyl9}.
Cells, which are initially all positioned in the source at $x=0$, migrate away from the full reservoir; as time advances, some of them move into the empty region on the right resulting in the advancing of  the front of the density profile  which gets closer to the sink at $x=L$. The effect of the sink is perceived only at large times, when cells arrive at the empty reservoir. Before that time, the slope of the density profile is strictly negative.  
The solution of  the ODEs for the density profile is in good agreement with the simulations; nevertheless, some differences are noticeable  
on the right part of the density profile  
where $\rho(x, t)\ll 1$ due to the closure approximations. 

\begin{figure}[h]
\includegraphics[width=0.35\textwidth, angle=-90]{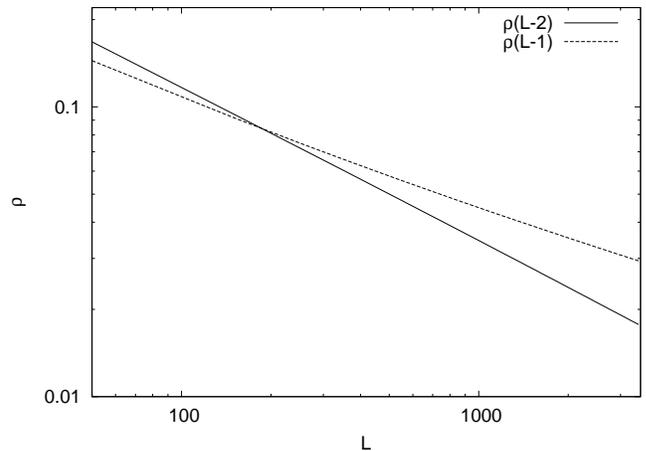}
\caption{\label{LongCyl}Gap junctional model on a cylindrical geometry with p=1. The straight line shows $\rho(L-1)$ and the dashed line represents $\rho(L-2)$. From a numerical fits $\rho(L-1)$ goes approximately as % 
$\frac{1}{L^{1/3}}$ and $\rho(L-2)$ goes approximately as 
$\frac{1}{{L^{1/2}}}$. }
\end{figure}

\textbf{Gap junctional model --- discontinuity.}
At large times, in the position at one lattice step away from the sink, the density  presents a discontinuity when $p=1$.
This is due to couples approaching the empty reservoirs. When they are in contact with it, the next favorable jump in the direction of the sink annihilates one cell of the couple, leaving the other in stall without functional gap junctional connections. It results in
a small accumulation of cells on the sites just before the empty reservoir. This phenomenon is quite evident in the stochastic simulations where the slope of the density profile between the sites $L-2$ and $L-1$ becomes positive. From the 
solution for the steady state equations of the gap junctional model, 
the discontinuity also appears when the source and the sinks are few lattice steps apart, for example $L=36$, 
but there is no inversion of the slope of the density profile. To obtain such results, it is necessary to increase the distance between the reservoirs. In fact, the density at a specific distance from the sink decreases when  the cylinder length  $L$ increases and the decrease is slower at position $L-1$ than at other distances; eventually, for $L$ large enough, it will hold true that $\rho(L-2)<\rho(L-1)$, Fig.~\ref {LongCyl}.
Hence, even though the strong effect of the closure approximations requires an increase of the distance between source and sink, taking into consideration the nearest neighbors  two-point connected correlation function in the equations for the evolution of the system is enough to analytically reproduce  the discontinuity when $p=1$. 
As   $p$ decreases, the discontinuity in the density profile at the last point close to the sink becomes  quickly less evident and at $p=0.99$, it disappears. The reason is that, in these cases, single cells can jump and do not accumulate; therefore, the slope of the density profile remains negative at all positions for all times, Fig.~\ref {GapDenCyl9}.

\textbf{Gap junctional model --- correlation.}
In Figs.~\ref {GapCorCyl1}, \ref {GapCorCyl9}, the  correlation obtained with the cellular automata  is very different from the  correlation from the solution of the ODEs. The effect of the approximations makes it impossible to recover the exact correlation values  of the stochastic  simulations, but 
one can see that the correlations from the analytical model   and simulations  share the same properties.        
The particular shape of the connected correlation function at a given time $t$, Figs.~\ref {GapCorCyl1}, \ref {GapCorCyl9}, divides the figure in three regions. In the \emph{first region} that goes from the source of the cells  to the point where the curve crosses the $x$ axis, the correlation has a negative tail, and this is due to the exclusion process  which forbids several cells from occupying the same site. This introduces in the system a short range repulsion between cells, especially close to the source where cells are crowded  resulting in a negative value of the correlation.     
The \emph{middle region} goes from where the correlation becomes strictly positive to the point  where the stochastically  simulated correlation is indistinguishable from zero (meaning  that there the error bar of the correlation  cross the $x$ axis). It
corresponds to the zone where the density starts to become low, and cells begin to feel the lack of neighbor cells. For any $p>0.5$, cells tend to maintain gap junctions with the neighborhood during their  motion due to the binding interaction, but the more the density approaches an approximative  value  
$1/3$, the harder it is to preserve a contact with other cells. As a result, cells  feel the crowding effect less, and the repulsive effect of the exclusion process is surpassed by the binding interaction. It is in this less dense region that, at the microscopical scale, it is possible to see couples moving together with a tendency not to separate until another neighbor gets close enough to form a new gap junction. At $p=1$, the effect becomes very strong, and cells moves only if they maintain at least one functional gap junction
and
as soon as the density drops down,  a population of moving gap junctional communicating couples and single stalled cells appears.  The \emph{third region} goes from the end of the previous region to the sink; here, the correlation is almost zero and clearly, the sink is yet too far from the population of cells to be perceived. 

Let us consider the time evolution of the connected correlation. At the beginning, the repulsion between the cells  is stronger, but with the advancing of time, it decrease. 
The peak of the correlation moves toward the sink with a velocity $v(t)=t^{-1/2}$ and its height slowly decreases. 
Eventually, the correlation front will reach the sink and its peak slowly disappears leaving a very long and small negative correlation tail. 
At very large time, when the correlation is at the steady state, the connected correlation between two nearest neighbor  sites parallel to the bases of the cylinder  shows some abrupt changes in sign in the proximity of the empty reservoirs. Indeed,  the large negative values of $C_\parallel(L-1)$ and  $C_\perp(L-2)$ indicate a small number of cell couples in perfect accordance with the explanation of accumulation of single cells  just before the sink.

Although  the results above rely only on simulations in the cylindrical and radial geometries with hexagonal tiling, we will argue in Sec.~(\ref{sec:level4})  that they are independent of the specific lattice and spatial dimensionality (in particular, the power that governs the decay of the velocity $v$ of the front is always $-1/2$).

\begin{figure}[h]
\begin{center}
\subfigure[\label{adhq02D} \hspace{2pt} ]{
\includegraphics[width=0.35\textwidth, angle=-90]{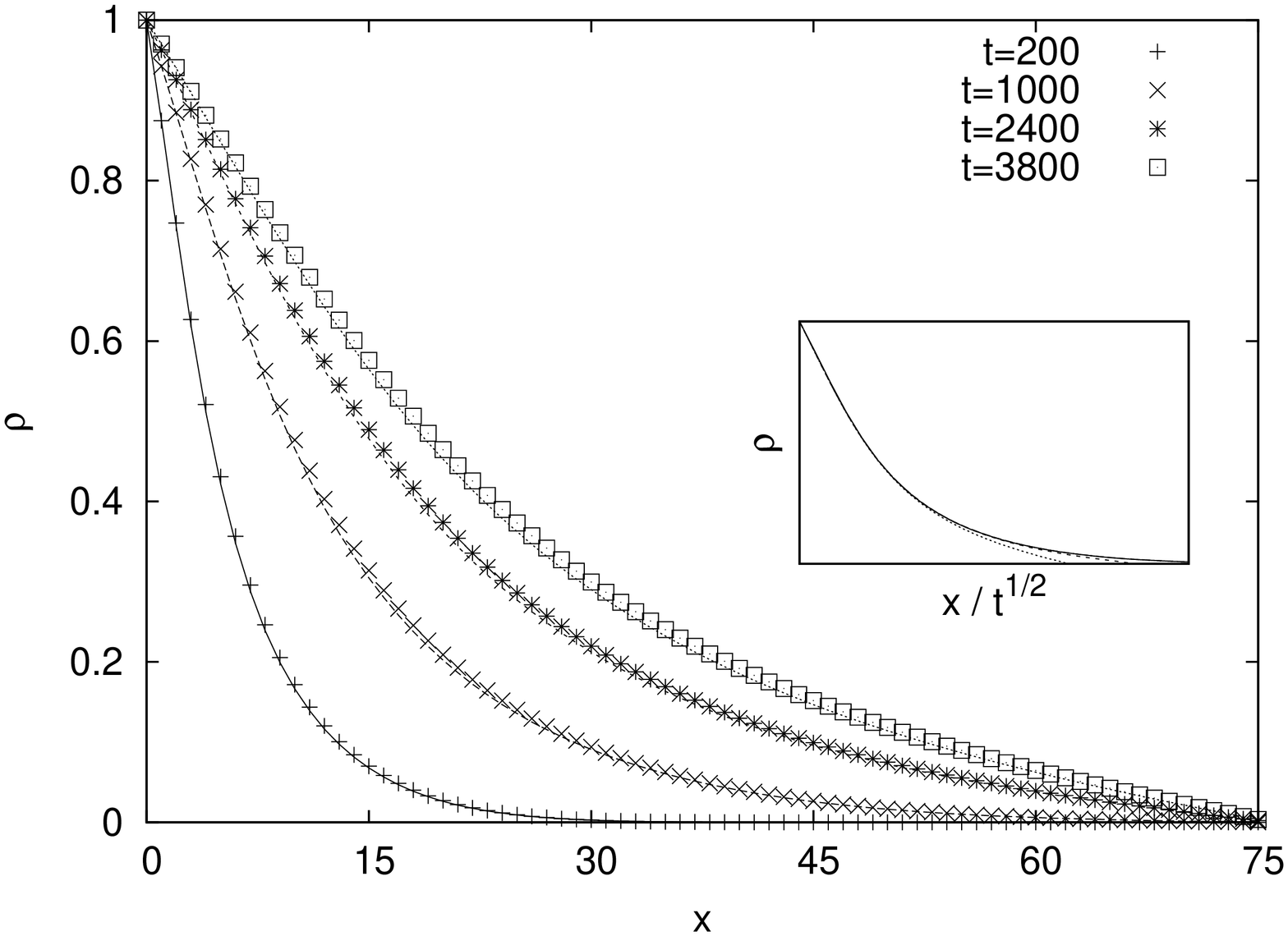}}
\vspace{6pt}

\subfigure[\label{adhq02C} \hspace{2pt}  ]{
\includegraphics[width=0.35\textwidth, angle=-90]{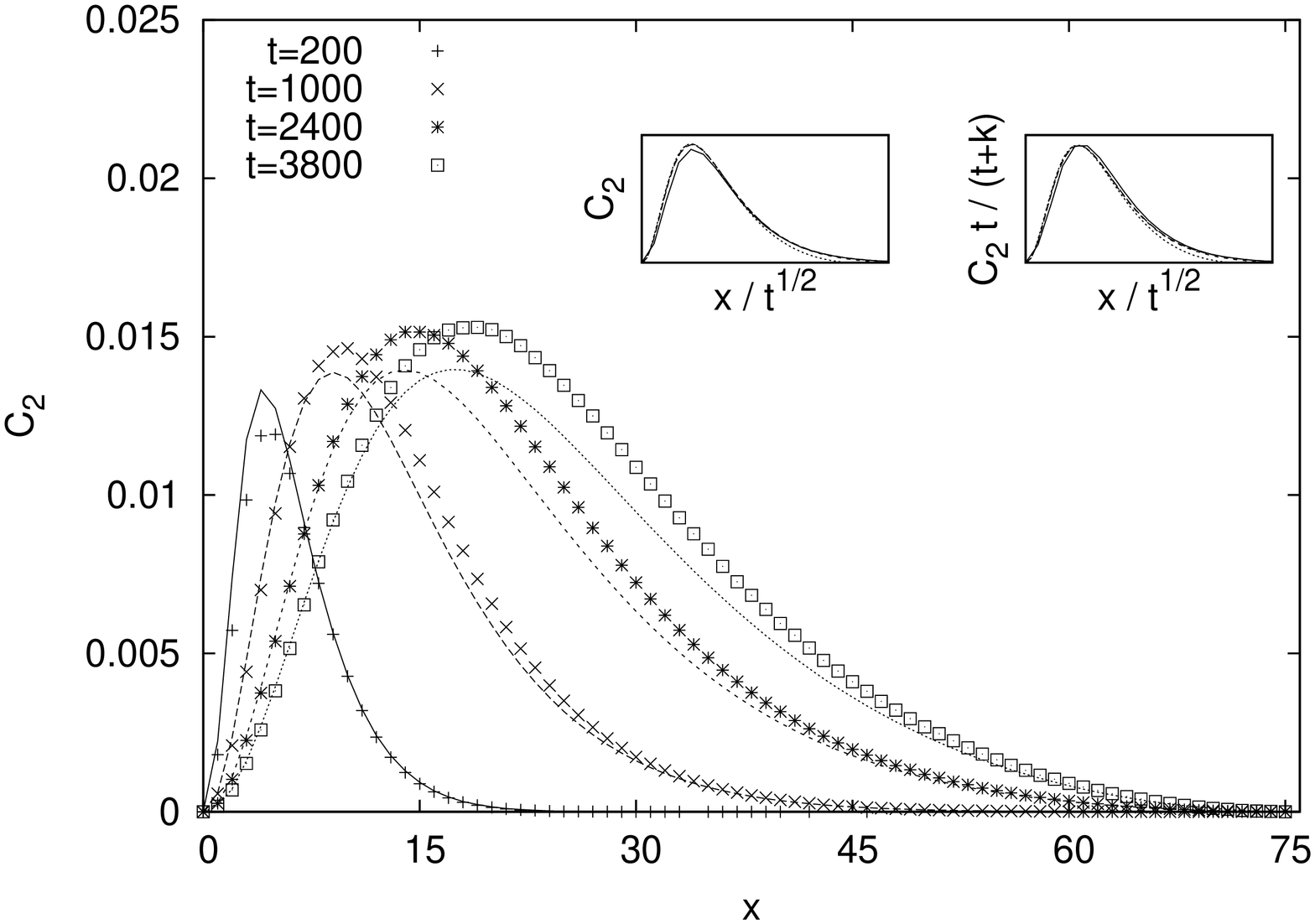}}
\vspace{6pt}
\caption{ \label{adhq02} Profile of \subref{adhq02D} the density and \Subref{adhq02C} the connected correlation at different times for the adhesion model on a cylindrical geometry with q=0.2 and L=75. The dots represent the results of the stochastic simulations and the error bars are smaller than the size of the symbols. The lines refer to the results of Eqs.~(\ref {rho}, \ref {C2def}, \ref {BC2}). The insets show the scaling behaviors. In \Subref{adhq02C}, the inset on the left-hand side refers to the analytical results, and that on the right shows the respective results obtained from cellular automata simulations. The parameter $k=6$ is properly chosen to show  the scaling of the stochastic simulations.}
\end{center}
\end{figure}

\begin{figure}[h]
\begin{center}
\subfigure[\label{GapRad_p1D} \hspace{2pt} ]{
\includegraphics[width=0.35\textwidth, angle=-90]{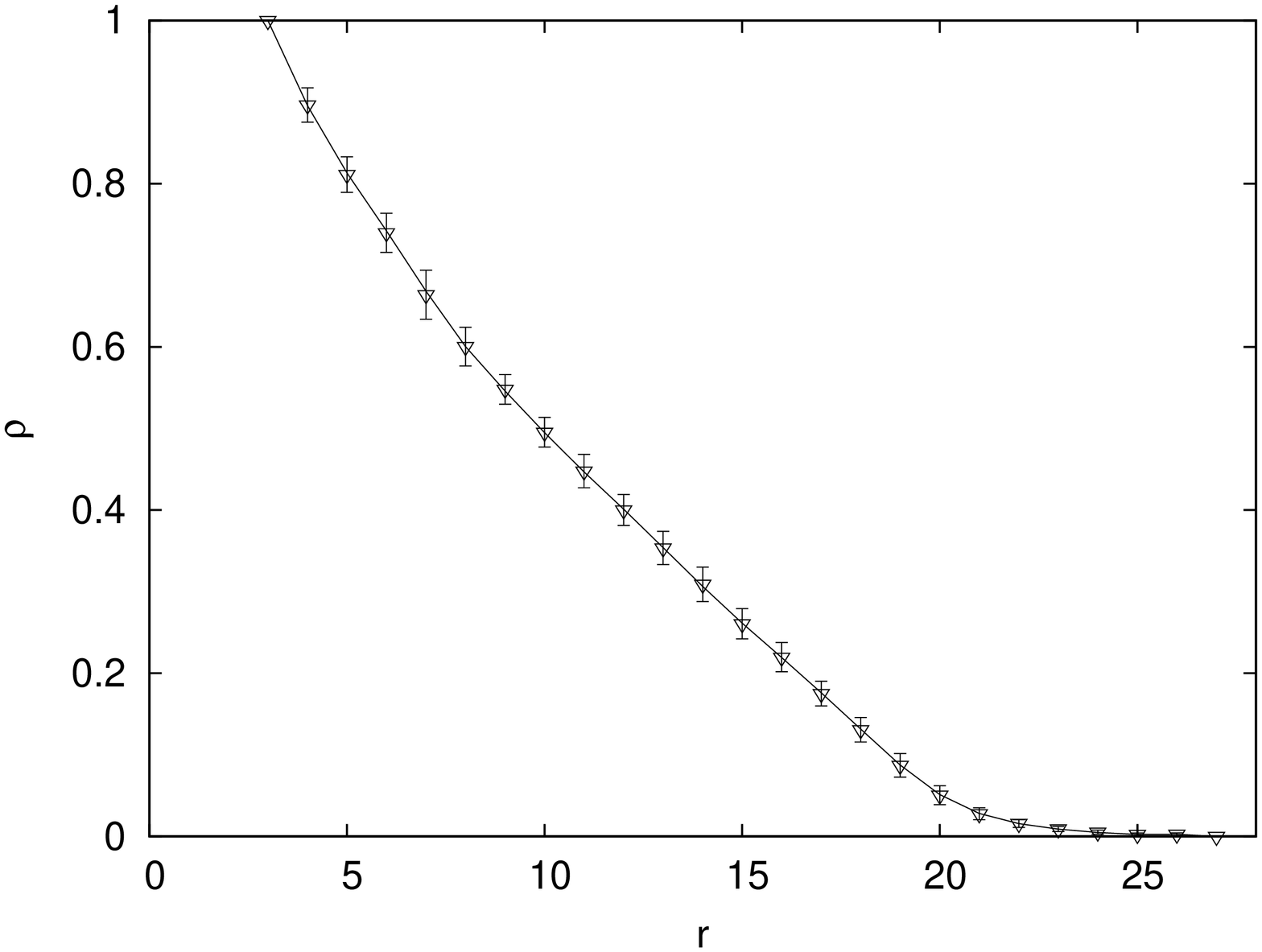}}
\vspace{6pt}

\subfigure[\label{GapRad_p1C} \hspace{2pt} ]{
\includegraphics[width=0.35\textwidth, angle=-90]{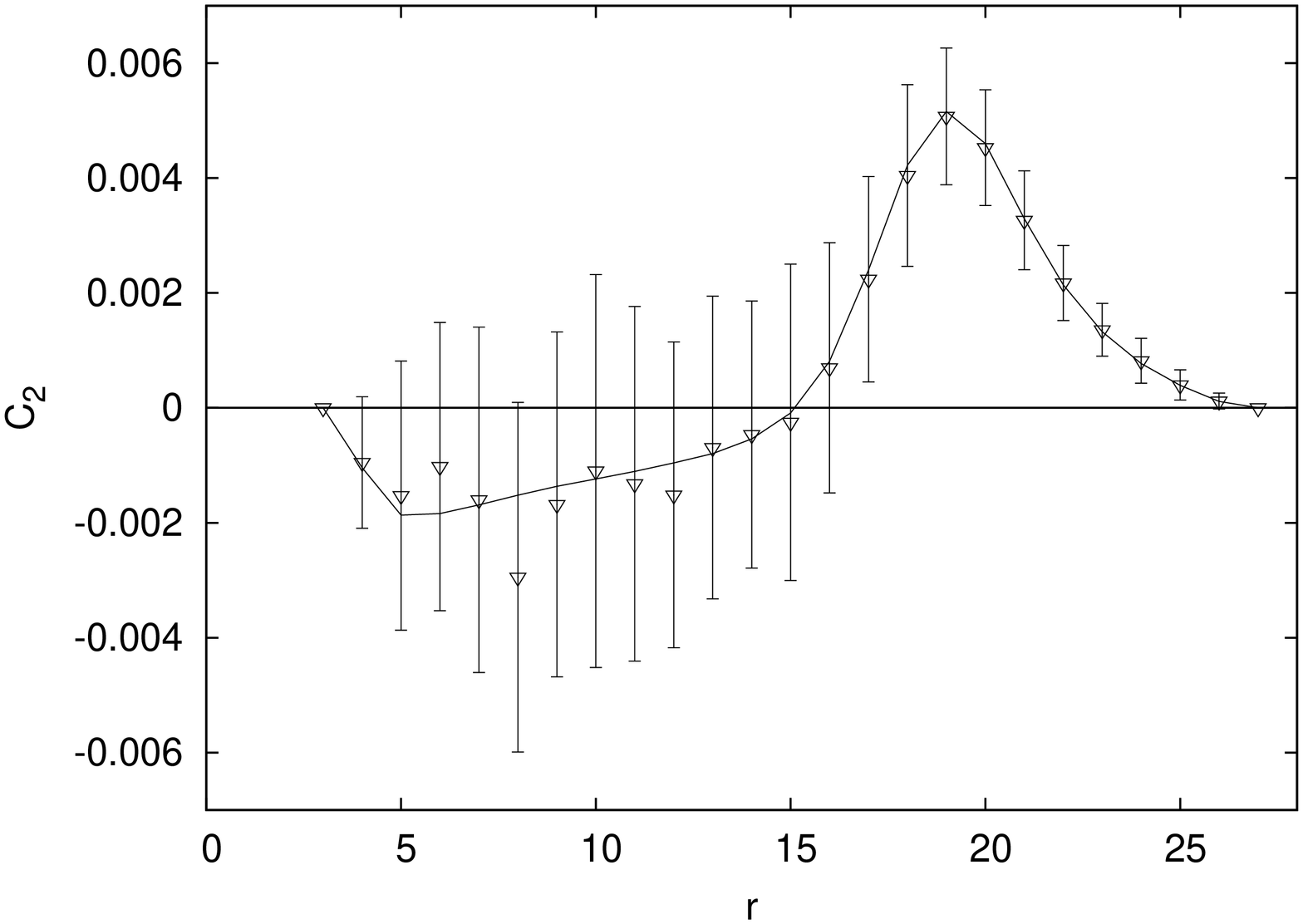}}
\vspace{6pt}
\caption{\label{GapRad_p1} Profile of \ref{GapRad_p1D}  the density  and \ref{GapRad_p1C}  the connected correlation at time $t=450$  for the gap junctional model on a radial geometry with $r_{src}=3$ and $r_{snk}=27$. The lines show the results of $3\cdot10^6$ averaged independent stochastic simulations at time $t$ with same initial conditions and parameter p=1. The dots and the bars represent the average and the error for only $10^3$ stochastic simulations.}
\end{center}
\end{figure}

\textbf{Linear model.}
The linear model with $\gamma=0$ does not differ qualitatively from the gap junctional model. The former has a smaller correlation than the latter, and for $\alpha=0$ at the steady state, the density presents a similar discontinuity near the sink. This model has the best agreement between the simulations and ODEs solutions.

\textbf{Adhesion model.}
For the adhesion model, the greatest differences between the values of the correlation obtained from the CA simulations and from the solutions of the ODEs  are in the region close to the source (see  Fig.~\ref {adhq02}), and they increase with the increasing of the parameter $q$. 
The repulsion produced by the exclusion process  is almost inexistent in the case of strong adhesion between cells, and the correlation is peaked at the position where the density is almost $1/2$
showing that, at the front of the density profile, cells are almost free, and they do not form any structures, but diffuse away. In contrast, at higher density, cells gather together, and because in part of the crowding and  of the high number of links, they get trapped, which produces strong correlations between cells.      
The correlation produced by the adhesion has a peaked shape which moves in the direction of the sink with a velocity  $v(t)\propto t^{-1/2}$. This is similar to what happens in the gap junctional model. On the other hand, its behavior is very different in the other two models. Very remarkable facts are that the height  of the peak is constant in time,  that it is much stronger than in the gap junctional case, and that, for large values of $q$, there is no negative correlation tail close to the source region from the beginning of the simulation on.
The same constant peak of the connected correlation can be obtained in the linear model with $\gamma<0$ and $\beta\leq0$.

The main differences between CA simulations and analytical models are due to the closure approximations which neglect the long range correlations (we have seen on Fig.~\ref{comparison} how taking short-range correlations into account improve the agreement with CA simulations). Observing the ratio between the values from the ODEs and from the stochastic simulations of the  two-point connected correlation function   at different times, which is approximately proportional to $\ln(t)$ and $\frac{t}{t+k}$ (depending on the contact interaction), we conclude that such a discrepancy is produced by the exclusion rule, a common factor between different models.     

\section{\label{sec:level4}The Hydrodynamic Limit}

In this section, we show and discuss how to derive, in the limit of 
infinitely large lattices, a finite set of coupled PDEs from the 
infinite system of ODEs of the previous section. Even if the PDEs are 
approximate, their solutions reproduce qualitatively the most probable 
(typical) behavior of the stochastic system at large times and large 
distances, the so-called hydrodynamic limit. In our case, this is not a 
very tight restriction since we have to disregard only a few time steps 
after the initial condition, and details on the scale of a single 
lattice step, such as the discontinuity close to the empty reservoir 
discussed earlier. On the other hand, the benefit is that only these 
PDEs will enable a discussion of the self-similar phenomena that take 
place in the different models \cite{barenblatt1996scaling}.

\subsection{\label{sec:level4a} Principle of the derivation}

We are interested in variations of the local cell density and local 
correlations on a length scale $R$ much larger than the lattice step 
$\lambda$. This length $R$ can be the distance between source and sink 
in a steady-state regime, or the size of the region of a Petri dish that 
cells exiting a spheroid have already invaded~Fig.~\ref {GeoRad}. If 
$\bs{r}_i \in \mathbb{R}^n$ is the position (in continuous space) of the 
center of lattice tile~$i$, we assume that there exists some function 
$\rho$ such that $\langle \eta_i \rangle = \rho(\bs{r}_i/R)$ for all $i$ 
and that $\rho$ is differentiable as many times as necessary w.r.t. its 
argument $\bs{r}_i/R$. This implies that the average number of cells in 
the lattice site $i$, $\langle \eta_i \rangle$, varies from the average 
number in the nearest neighbor site $j \in \mathcal{V}(i)$ from no more 
than a quantity of the order of $1/R$. On the light of the previous 
results, we know that it is an excellent approximation most of the time. 
The exceptions are in the set-up with migration out of a spheroid, at 
short times after the initial condition when $\rho$ has a steep slope 
(but this lasts less than $\approx$~10 time steps), and in the steady 
state close to a sink of cells, for some particular values of the 
parameters of the models.

Likewise, we assume that there exists a function $C_2$ such that the 
connected correlation function between nearest neighbor sites $i$ and $j 
\in \mathcal{V}(i)$ reads $$\langle\eta_i\eta_j\rangle - \langle \eta_i 
\rangle \langle \eta_j \rangle = R^{-\xi} C_2 
\left(\frac{\bs{r}_i+\bs{r}_j}{2R},\bs{r}_j-\bs{r}_i \right).$$ Since 
the correlation between the sites $i$ and $j$ does not depend on the 
order of $i$ and $j$, it is natural to use the middle point of $i$ and 
$j$ as the first argument of $C_2$. The second argument of $C_2$ can 
only take a finite number of values since the lattice is regular and it 
gives the direction of the two-point connected correlation function 
(sign is unimportant, $i$ and $j$ can be exchanged). To make expressions 
shorter, in the sequel, we will write $C_2 \left( \bs{r}, \bs{r}_j-\bs{r}_i \right)$ as $C^l$ where $l$ is the number of the 
direction $\bs{r}_j-\bs{r}_i$. Finally, in some of the 
previous simulations, we have seen that finite regions of the system may tend to 
homogenize (the density $\rho$ gets uniform). Therefore, we introduce the 
possibility that $C^l$ vanishes as $R$ goes to infinity, and we assume 
that it involves a power law with the exponent $\xi$ to be chosen later. The exponent $\xi$ must be nonnegative since the correlation function is always between -1 and 1.

It is generally not possible to define a regular function of two 
arguments, say $C_2(\bs{r}_i/R, \bs{r}_j/R)$: indeed, $C_2(\bs{r}_i/R, 
\bs{r}_{j'}/R)$ may have a difference of order 1 (not $1/R$), if $j'$ is 
another nearest neighbor of $i$ than $j$, but without 
$\bs{r}_{j'}-\bs{r}_i$ being colinear to $\bs{r}_j-\bs{r}_i$. This is 
for instance the case on the cylinder of Fig.~\ref {geoF} where $C^1$ and 
$C^2$ (e.g. correlations along the cylinder axis and perpendicular to 
the cylinder axis) can be quite different one from another because the 
boundary conditions break the rotation invariance of the lattice. The 
same holds if $j'$ is not a nearest neighbor of $i$.

Then we insert the expressions for $\rho$ and $C_2$ defined above into 
the set of discrete equations~(\ref{sys}) for the time evolution of 
$\rho(i)$ and $C_2(i, j)$, and we perform a Taylor expansion in powers 
of $R^{-1}$ on the right-hand side of the equations around the points 
$\frac{\bs{r}_i}{R}$ and $\frac{\bs{r}_i+\bs{r}_j}{2R}$ respectively, 
neglecting terms of orders three and above in $R^{-1}$. There are $1+d$ 
equations for each lattice site, where $d$ is the number of different 
directions, hence a total of $(1+d)N$ equations for the whole lattice of 
$N$~sites. But, since the equations are independent of the site up to an 
index shift, except for the sites on the boundaries of the system, we 
end up with only $1+d$ PDE, plus boundary conditions --- the 
(approximate) ``continuous space model'', or ``hydrodynamic limit''.

 Of course, one could systematically generalize this procedure to take 
into account correlations at distances larger than one lattice step or 
between three or more points --- this may further improve the quality of 
the approximation of the continuous space model. To do this, one can 
redefine the multi-points density function: 
$$\rho_p(\eta(\bs{r}_1),\ldots,\eta(\bs{r}_p))= 
\rho_p(\eta(\Delta\bs{r}_1),\ldots,\eta(\Delta\bs{r} _p); \bs{r}_C)$$ 
where $\bs{r}_C$ is the centroid $\frac{\sum_k^p \bs{r}_k}{p}$, and 
$\Delta\bs{r}_k=\bs{r}_k-\bs{r}_C$ is the position of k-th point of 
$\rho_p$ with respect to the centroid. Under the assumptions above, 
$\rho_p$ with fixed $\Delta\bs{r}_k$ for all $k$ is a regular function 
of $\bs{r}_C$. Then we could perform a series expansion, at fixed 
$\Delta\bs{r}_k$, in powers of $1/R$, of the multipoint probability 
density which can be in turn inserted into the system of discrete 
equations to yield a system of PDE. But here, we restrict ourselves to equations 
with $\rho$ and $C^j$ only.

\subsection{Explicit expressions}

Before explicitly applying  the Taylor expansion to the discrete equations \eqref{sys} for the evolution of the density and the connected correlations, it is important to remark that microscopic symmetries related to the regularity of the lattice
reflect on the hydrodynamic limit. 

Therefore, the  equation of the two-point correlation between neighbor sites in one direction can be changed in
the equation for the correlation in
another direction by applying rotations of $\pi/3$ rad as consequence of the invariance of the  hexagonal lattice under such transformations. This symmetry holds  everywhere in the region of interest between the reservoirs, except in the sources and in the sinks.

To easily express the results and the symmetries of the system, we introduce a set $D$ of unitary vectors identifying the three main lattice directions 
$e_0, e_1,  e_2, \in\mathbb{R}^2$ such that:

\begin{eqnarray}\label{Dvector}
\begin{array}{lll}
$$e_0=
\begin{pmatrix}
0\\
1
\end{pmatrix},\quad  $$&
$$e_1=
\begin{pmatrix}
\frac{\sqrt{3}}{2}\\[0.3em]
\frac{1}{2}
\end{pmatrix},$$&
$$e_2=
\begin{pmatrix}
-\frac{\sqrt{3}}{2}\\[0.3em]
\frac{1}{2}
\end{pmatrix},$$
\end{array}
\end{eqnarray}
and the directional derivative $\nabla_{i} = e_i.\nabla$, where the index $i$ means that the derivative is taken along the direction $e_i$. 

In the case of the linear model, after performing the expansion in series in respect to the lattice distance $\lambda$ to each term of the right-hand side of Eq.~(\ref {sys}), the results are:
\begin{equation}\label{rhoCont}\partial_t \rho(\bs{r}, t)=\!\!\!\sum_{\!\!\!\!\!\!k\in\{\alpha,\beta,\gamma\}}\!\!\!\!k\left[\frac{B_{k,\rho}}{R^2}+\sum_{j=1}^3\frac{B_{k, C^j}+B_{k,\rho, C^j} }{R^{2+\xi}}\right], 
\end{equation}

\begin{equation}\label{C2Cont}
\frac{\partial_t C^j(\bs{r}, t)}{R^\xi}= 
\!\!\!\!\sum_{\!\!\!\!\!\!k\in\{\alpha,\beta,\gamma\}}\!\!\!\!k\left[ \frac{A_k^j}{R^\xi} + \frac{B_{k,\rho}^j}{R^2}+\!\sum_{i=1}^3 \frac{B_{k, C^i}^j+B_{k,\rho, C^i}^j}{R^{2+\xi}}\right], 
\end{equation}
where the terms named $A$ are the zero order  and the terms named B are the second order of the Taylor expansion.
In both equations, for  the density and the correlations, there are no terms with odd order due to reflection symmetries of the lattice with respect to the lattice main directions.
The equation for the density does not have any zero order term because the number of cells is locally conserved.  
The terms $B_{k,\rho}$, and $B_{k, C^i}$ both contain terms proportional to the
parameter $k\in\{\alpha, \beta, \gamma\}$ and each term only depends  on the density and the correlation in direction $i$, respectively. $B_{k,\rho, C^i}$  is the interaction part that takes into account the coupling between the density and the correlation in direction $i$. The same definitions hold for the terms in Eq.~(\ref {C2Cont}) where the upper index  $i$ (resp. $j$)  refers to one of the main directions of the two-point connected correlation. From here on, all upper indices must be considered modulus three, even though it is not explicitly written.
The details and expressions of all terms of the Taylor expansion $A$ and $B$ are in the Appendix.

Defining  $S$ and $s$ the spatial subsets of $\mathbb{R}^2$  where there are  sources and sinks respectively, the border and the initial conditions used to determine the solution of the system of partial differential equations are: 
\begin{equation}\label{ICcont}
\begin{array}{ll}
\rho(\bs{r}, t)=1 & \forall\bs{r}\in S,\nonumber\\
\rho(\bs{r}, t)=0 & \forall\bs{r}\in s,\nonumber\\
C^i(\bs{r}, t)=0 & \forall\bs{r}\in s\cup S,\nonumber\\
\rho(\bs{r}, t=0)=C^i(\bs{r}, t=0)=0 & \forall\bs{r}.\nonumber
\end{array}
\end{equation}

The computation of the hydrodynamic limit for  the adhesion model and the gap junctional model do not present  ulterior problems, but a subtle more analytical complexity, and it can be performed  in the same way as for the linear model. The hydrodynamic limit introduces some approximations in comparison to the discrete equations  because one disregards  terms of order bigger than $1/R^2$ in the series expansion. The numerical solutions of the PDEs are obtained by first re-discretizing the space derivatives using a second order finite difference  method and then following the same numerical procedures as for the ODEs. They are accurately in agreement with the solutions of the ODEs for all the models proposed except when close to the sink.
The analytical solutions are $C^{(\infty)}$; therefore, at the last step and two lattice steps away from the empty reservoirs for the density and the correlations respectively,   the numerical solutions  smoothly go to zero in all cases
making it impossible to retrieve the discontinuity observed in the discrete representation for some values of the parameters of the models  at the steady state.   

\subsection{\label{sec:level4c} Self-similar behaviors}
When non-interacting agents randomly jump to the nearest neighbor site with a constant transition rate,  in the continuous  limit, the  concentration $\rho$ of agents at the position $\bs{r}$ obeys: $\partial_t\rho(\bs{r}, t)=D\Delta\rho(\bs{r}, t)$. 
A generalization of the previous equation is the diffusion equation in porous media \cite{simpson-pme}: 
$\partial_t\rho(\bs{r}, t)=\bs{\nabla}[D(\rho)\bs{\nabla}\rho(\bs{r}, t)]$ which is useful to describe the concentration of agents when the agents interact \cite{witelski-pme-interaction}.   
Both equations with the initial condition $\rho(\bs{r},t\!=\!0)=\delta(\bs{r})$ show a self-similar behavior. 
Indeed, the solution of the diffusion equation at a given time $t_1$ is self-similar to the solution at time $t_2$: $\rho(\bs{r}, t_1)=\rho(\bs{r}\sqrt{\frac{t_1}{t_2}}, t_2)$. In this specific case, the self-similarity of the density function is exact, but in other cases, when correlations are taken into consideration, such behavior holds true only approximately and at large times far from the initial transient regime \cite{sachdev-selfsimilarity-asymptotic}. 

This form of self-similarity does not depend on the specific lattice, or space dimension in which the lattice is embedded. Indeed, 
if we choose a different type of tiling or a different space dimension, the diffusion equation above will keep the same structure, with a first-order time derivative and a (sum of) second-order space derivatives. Only numerical coefficients in $D$ will change.

%This form of self similarity does not depend on the specific lattice, or space dimensions in which the lattice is embedded. Indeed, if we choose a different type of tiling and we change properly the concept of nearest neighbor such that the rules  do not change and the cells are allowed to jump, the equations     

For the geometrical  dispositions of the reservoirs proposed in this work, the presence of a sink destroys the self-similar behaviors of the density of cells; therefore, 
we consider the situation where the sink is sufficiently far ($T\leq L$, where $T$ is the maximum time we run the stochastic simulation) so that the cells can migrate for long enough without perceiving the sink.

To investigate the self-similarity of both the density and the two-point connected correlation, we exploit the PDEs  derived from the hydrodynamic limit for the linear model to retrieve the scaling of the solutions.
When there are no correlations, the Eq.~(\ref {rhoCont}) becomes a particular type of diffusion equation in porous media; therefore, after eliminating the factor $R^{-2}$ thanks to the change of variable $t=R^2 \, \tilde{t}$, we can write the solution of  Eq.~(\ref {rhoCont}) as: 
\begin{equation}\label{iter_rho}
\rho(\bs{r}, \tilde{t})=f(\frac{\bs{r}}{\sqrt{\tilde{t}}})+\epsilon g(\bs{r}, \tilde{t}) 
\end{equation}
where the function $g(\bs{r}, \tilde{t})$ on the right-hand side tells us how the exact solution of the density profile differs from the self-similar result $f(\frac{\bs{r}}{\sqrt{\tilde{t}}})$ obtained in case of no correlations. 
The parameter $\epsilon$ is the magnitude of such discrepancy and for  negligible correlations in comparison with the density, it holds true that $\epsilon\ll1$. 
Inserting the solution $f(\frac{\bs{r}}{\sqrt{\tilde{t}}})$ into  Eq.~(\ref {C2Cont}), one can find the approximate solutions for the correlations and use them again in the equations for the density. Repeating the process iteratively,  more accurate solutions for the density and the correlations can be found, and at each repetition, it is possible to check the validity of the hypothesis $\epsilon\ll1$.   
In the insets of Figs.~\ref {GapDenCyl1}, \ref {GapDenCyl9}, \ref {adhq02D}, the stochastic simulations  of each model  at different times are scaled 
using the scaling behaviors of the function $f(\frac{\bs{r}}{\sqrt{t}})$.
We have observed that the same self-similar behavior holds for stochastic simulations of the linear model with various values  of parameters which reproduce the adhesion or gap junctional model behaviors.   
The perfect overlap between the curves shows that analyzing the scaling  of $\rho(\bs{r}, t)$ under the conditions  of negligible correlations is very good. As we stated above, this self-similar behavior will hold also in different space dimensions and on other lattices because the structure of Eq.~(\ref {rhoCont}) will be the same (first-order time derivative, second-order space derivative), even though numerical coefficients may be different from those provided in the Appendix.

In the sequel, we change the notation $\tilde{t}$ to $t$ for simplicity, and we consider that $\rho(\bs{r}, t)=f(\frac{\bs{r}}{\sqrt{t}})$. Eq.~(\ref {C2Cont}) now reads:
\begin{equation}\label{C2Contasympt}
\frac{\partial_t C^j(\bs{r}, t)}{R^{\xi+2}}= 
\!\!\!\!\sum_{\!\!\!\!\!\!k\in\{\alpha,\beta,\gamma\}}\!\!\!\!k\left[ \frac{A_k^j}{R^\xi} + \frac{B_{k,\rho}^j}{R^2}+\!\sum_{i=1}^3 \frac{B_{k, C^i}^j+B_{k,\rho, C^i}^j}{R^{2+\xi}}\right].
\end{equation}

\textbf{Linear model --- only exclusion interactions.}
To begin, let us consider the simplest case $\alpha\neq0$ and $\beta=\gamma=0$.
In Eq.~(\ref {C2Contasympt}), choosing $\xi \ne 2$ leads either to a trivial solution $C=0$ or to an inconsistent  solution $\nabla \rho=0$. Therefore, the only possible choice is $\xi=2$ which results in $C^j(\bs{r}, t)\sim\Delta\rho(\bs{r}, t)\sim\frac{h(\frac{\bs{r}}{\sqrt{t}})}{t}$, while $\partial_t C^j(\bs{r}, t)\sim\nabla_j^2 C^j(\bs{r}, t)$ are asymptotically negligible. 

\textbf{Linear model --- mimicking gap junctions $\bf{p<1}$.}
The same discussion  is valid for $\gamma=0$ and both $\alpha$ and $\beta$ different from zero. This brings us to 
the same self-similar behavior $C^j(\bs{r}, t)\sim\Delta\rho(\bs{r}, t)\sim\frac{h(\frac{\bs{r}}{\sqrt{t}})}{t}$. In the left insets of Fig.~\ref {GapCorCyl9}, the numerical solutions of $C_\parallel$ for the gap junctional model are rescaled using  $C^j(\bs{r}, t)=\frac{h(\frac{\bs{r}}{\sqrt{t}})}{t}$ as self-similar behaviors. The good agreement given by the overlapping of the curves also shows that the gap junctional model for $p<1$ shares the same self-similar behaviors as the linear model.
 
The right inset of  Fig.~\ref {GapCorCyl9} shows that the results of the stochastic simulations scale as $C^j(\bs{r}, t)=\ln(t) \frac{h(\frac{\bs{r}}{\sqrt{t}})}{t}$. The logarithmic correction in the scaling of the correlations  is due to the long range correlation produced by the exclusion process, and they cannot be retrieved using the hydrodynamic limit equations.

\textbf{Linear model --- mimicking gap junctions $\bf{p=1}$.}
The case $\alpha=\gamma=0$ and $\beta\neq0$ is particularly interesting because  the solution  of Eq.~(\ref {rhoCont}) with no correlations is no longer a smooth function of the position, but it is a piecewise function with two regions: the part closer to the source, where the density  decreases with a finite negative slope as its distance from the full reservoir increases,  and the part that goes  from the positions where the density becomes zero onwards,  in which the density is uniformly null \cite{pme-selfsimilarity}. 
  
The discontinuity in the first derivative of the density profile for $\alpha=0$ is true only when there are no correlations, while, when correlations are introduced, such discontinuity disappears.
Hence, always under  the condition $\rho(\bs{r}, t)=f(\frac{\bs{r}}{\sqrt{t}})$, in the region where $\rho(\bs{r}, t)\neq0$, the correlation rescale as $C^j(\bs{r}, t)= \frac{h(\frac{\bs{r}}{\sqrt{t}})}{t}$.

 On the other hand, in the region where the density is zero, all terms containing $\rho$ or its derivatives are zero, and Eq.~(\ref {C2Contasympt}) becomes:
\begin{equation}\label{C2Contrhonull}
\frac{\partial_t C^j(\bs{r}, t)}{R^{\xi+2}}= 
\beta\left[ \frac{A_\beta^j}{R^\xi} + \!\sum_{i=1}^3 \frac{B_{\beta, C^i}^j}{R^{2+\xi}}\right].
\end{equation}

The leading order term in $1/R$ must be zero. Hence, $A^0_\beta=A^1_\beta=A^2_\beta=0$ and Eq.~(\ref {C2Contrhonull}) becomes a system of three standard diffusion equations without sinks or sources. From the expression of $A^j_\beta$, we conclude that $C^0=C^1=C^2$. Therefore, Eq.~(\ref {C2Contrhonull}) simplifies to a standard diffusion equation for any of the quantities $C^j$. If we look for a nontrivial result for $C^j$, the three diffusion equations become identical to: $\partial_t C^0 = \frac{1}{24} \Delta C^0$.

 From the conservation of the correlation $C^j$ and the solution of the diffusion equation follow that the correlations in all directions scale as: $C^j(\bs{r},t)=\frac{q(\frac{\bs{r}}{\sqrt{t}})}{\sqrt{t}}$. If the complete solution  Eq.~(\ref {rhoCont}) is used instead of the approximate solution of the density with no correlations, then the region with $\rho=0$ changes in $\rho\ll1$. This correction,  even though it is very small ($\epsilon\ll1$), solves the inconsistency of having a correlation different from zero in a region with no cells, and leaves unchanged the self-similar behavior. 
 
In the top and bottom left insets of Fig.~\ref {GapCorCyl1},  we see that the numerical solutions of the ODEs at different times for the gap junctional model with $p=1$ reproduce the self-similarity  of the case $\alpha=\gamma=0$, and $\beta\neq0$, where the negative tail of the correlations scales as $\frac{h(\frac{\bs{r}}{\sqrt{t}})}{t}$ and the positive peak of the correlations as $\frac{q(\frac{\bs{r}}{\sqrt{t}})}{\sqrt{t}}$. The top and bottom right insets of Fig.~\ref {GapCorCyl1} show the scaling for the stochastic results of the gap junctional model with $p=1$, and also in this case the logarithmic correction is due to the long range correlations.

\textbf{Linear model --- mimicking adhesion.}
In the case with the parameter $\gamma \neq 0$, the scaling behavior is given  by $A_\gamma^j$ which contains terms proportional to positive powers of $\rho(\bs{r}, t)$ and $R^\xi$. One has to choose $\xi=0$ to get consistent results. Consequently, the self-similar form of the correlation is $C^j(\bs{r}, t)\sim\rho(\bs{r}, t)\sim f(\frac{\bs{r}}{\sqrt{t}})$, meaning that the height of the correlation peak remains constant during the migration process. 
When both $\beta $ and $\gamma$ are negative, the linear model mimics the behaviors of the adhesion model. Indeed, scaling the numerical solutions at different times of the ODEs for the adhesion model with the same scaling behavior of the corresponding linear model, we have a perfect overlapping of the curves, left inset of Fig.~\ref {adhq02C}. In the right inset, the results of the stochastic simulations show an increasing of the height of the peak due to long range correlations. In the adhesion model such effects are much smaller and asymptotically negligible  than in the other cases where correlations produce  logarithmic correction.
In all the insets in Figs.~\ref {GapCyl1}, \ref {GapCyl9}, \ref {adhq02}, the curves that deviate the most from the self-similar behaviors are those where the sink has been reached and the constraint $t \leq L$ does not hold any more.

\section{\label{sec:level5}Discussion}
\subsection{\label{subsec:level:1} Measuring the correlations in 
experiments}

Let us try to estimate for what parameters (population size, number of 
repetitions, duration of experiment) one could get an exploitable 
measure of the connected correlation function in migration assays of 
cells on Petri dishes. To be closer to this experimental situation, we 
performed stochastic simulations and got numeric solutions of Eqs.~(\ref {sys}, \ref {BC}) on 
the radial geometry of Fig.~\ref {GeoRad}. 
In this geometry, the density profile is computed by averaging the densities over the sites belonging to the same annulus centered on the source center, $O_{src}$, and one lattice step thick, so that the density depends only on the distance from the source center. The profile of the nearest neighbor two-point connected correlation is obtained by averaging the correlation over the six nearest neighbors of one of its sites, then the result is averaged over all the sites belonging to the respective annulus as for the density profile. 
At large times, the radius of the outer region invaded by cells is large, so 
that one can expect that values of the observables in that region will 
be similar to what is found with the cylindrical geometry of 
Fig.~\ref {GeoCyl}. It may be different both at short times and close 
to the center of the disk. 
However, it turns out that the ``correlation 
wave'' observed and discussed in Sec.~(\ref {sec:level3}) and 
Sec.~(\ref {sec:level4}) is still present in this setup and even amplified.
In addition, in the case of the gap junctional model where correlations are the most 
important far from the center of the disk, the relative error on the 
measure of the correlations tends to decrease with time, since the 
measure is done by averaging over a region which gets larger and larger.
To show these results, in Fig.~\ref {GapRad_p1}, we have plotted the  average of 1000 repeated simulations at a given time and  the respective error bars in comparison with the results obtained from the average of a much higher number of  simulations for the gap junctional model.  The figure, on one side, shows that it is not necessary to have a prohibitively high statistic to observe the two-point connected correlations, while, on the other side, it shows that simulations on the radial geometry are very useful to compare the proposed models with experimental data on Petri dishes.

\subsection{\label{subsec:level:2} Perspectives}

 Let us comment on the connection of the present approach to other 
techniques. The structure of Eq.~(\ref {iter_rho}) can be seen as an iterative 
computation of the local density $\rho$, where the first step would 
yield the mean-field approximation to the evolution equation of $\rho$ 
(all correlations $C_2$ being replaced with 0) and the second step 
involves an improvement to this equation that is expressed by means of $C_2$. 
If one formally integrates the evolution equation for $C_2$, replacing 
in it the value of $\rho$ that solves the mean-field equation for 
$\rho$, $C_2(t)$ gets expressed as an integral of $\rho(t)$ over the 
time range $0 \le t' \le t$. Then, replacing $C_2(t)$ in the equation 
for $\rho$ yields an integro-differential equation for $\rho(t)$ --- a 
non-markovian model with memory 
kernels~\cite{deroulers-monasson-proccont}. This is much like what is 
obtained in the Mori-Zwanzig formalism~\cite{evans-morris-livre}, the 
non-markovian character being due to a tentative of description of a 
complex situation (many spatially inhomogeneous configurations) with 
a simple quantity (the local density $\rho$) by a kind of 
``projection''. However, there are many ways to do this systematic 
reduction of numbers of freedom~\cite{evans-morris-livre, doi, 
taueber-et-al-revue-rg-pour-reaction-diffusion, 
deroulers-monasson-proccont}, and, when truncated to the first or second 
step of iteration, not all schemes yield analytical approximation with 
the same degree of agreement to simulations.

 It would also be interesting to have a deeper understanding of which 
situations (and possibly a criterion to discern them) allow one to get 
analytical results of good quality while taking into account only 
short-range correlations. Our situation might be related to the case of 
the Smoluchowski theory of aggregation discussed 
in~\cite{taueber-et-al-revue-rg-pour-reaction-diffusion}. The success of 
this method in spite of the simple approximation it relies on may be 
explained, at least for some of the models it was applied to, by the 
observation that there is no propagator renormalization in the 
corresponding field theories (in an RG approach), and hence no anomalous 
dimension for the diffusion constant or the fields.

 There are natural and biologically relevant extensions of the models 
considered in this work that can be studied with the same techniques, 
and we plan to deal with some of them in our future work. First, one can 
incorporate the effect of cell proliferation. According to previous 
works~\cite{simpson-baker-correcting-mean-field-uniform-in-space, 
baker-simpson-correcting-mean-field-spatially-dependent, 
simpson-baker-correcting-mean-field-adhesion-1d}, cell proliferation 
leads to short range correlations (because daughter cells are close to 
the position of the mother cell) which have the same order of magnitude 
as the ones produced by our contact interactions. In a realistic model, 
one should probably take care of the migration-proliferation dichotomy 
in the behavior of cells~\cite{giese-invasion, 
fedotov-et-al-migration-proliferation-dichotomy-anomalous-switching},  
which may enhance correlations. Then, the techniques can be 
useful for other models of exclusion processes with interactions like 
the ones of~\cite{simpson-et-al-mesoscale-patterns} 
and~\cite{landmann-fernando-myopic-random-walkers}. It can also 
straightforwardly be extended to 3D migration, for which experimental 
data is more difficult to obtain, but still accessible thanks e.g.\xspace 
to confocal microscopy. The techniques of our work can be used in spaces with dimension other than two, or on other lattices than the hexagonal tiling. In these cases, as explained in Sec.~(\ref{sec:level4c}), we would find an unchanged self-similar behavior for each respective model.

 An extension to disordered systems would also be useful. In the simple situation of one or a few obstacles or heterogeneities in the lattice with a known position, or a boundary with an irregular, but known shape, our techniques can be straightforwardly extended by writing equations specific to this geometry. The invariance and symmetry properties of the density and correlation functions are lost, but we expect the same large-time self-similar behaviors. A more complicated situation with many obstacles or inhomogeneous space can be modeled for instance by deformed 
lattices~\cite{drasdo-et-al-classification-croissance-prl, 
aubert-et-al-migration-sans-astrocytes-sains}. These cases may yield more 
realistic models of biological processes and make it possible to avoid some artifacts caused by regular lattices. As a first step to address this problem, one can use the so-called annealed approximation, where the computation of the density of cells is performed as if the obstacles would be redistributed randomly at each time step. The result is an approximate value for the diffusivity $D$ that depends on the density of obstacles, valid for not too large densities of obstacles~\cite{effet-inhibition-jonctions-communicantes}. In this case, we expect that the self-similar behaviors that we found in this work still hold.

\subsection{\label{subsec:level:3} Conclusion}

 We have shown how to extend simple mean-field analytical approximations 
of spatially inhomogeneous exclusion processes with local contact-like 
interactions as can be found in biological situations of interacting 
migrating cells. Supplementing the local density of agents (or cells) 
with short-range correlations in the analytical, deterministic, 
macroscopic approximations of the stochastic processes yields not only 
results in better agreement with stochastic simulations. It can also be 
used to infer more precise estimates of the interaction parameters from 
experimental data about density and correlations of migrating cells than 
from density values alone, when the type of cell-cell contact 
interactions is known. If the latter is unknown, it can be discovered by 
comparison with the classification scheme that we provide, which was 
made possible thanks to the PDEs approximation 
of the exclusion processes: the self-similar 
behavior of the two-point connected correlation function is highly dependent on the type 
of contact interactions, adhesive or contact-maintaining for instance, 
at the microscopic level.

\begin{acknowledgments}
The group belongs to the CNRS consortium CellTiss. CD acknowledges the 
hospitality of the KITP, where part of this work was done. We 
acknowledge financial support from the Comit\'e de financement des 
th\'eoriciens of the IN2P3.
\end{acknowledgments}

\appendix*
\section{Hydrodynamic limit explicit expressions}

Using the techniques developed  in Sec.~(\ref{sec:level4a} ),  we explicitly derived the $A$ and $B$ terms introduced  in Eqs.~(\ref {rhoCont}, \ref {C2Cont}) for the hexagonal tiling in the cases of gap junctional, adhesion, and linear models. For the sake of shortness, we write only the expressions of the linear model, which are enough to study all the self similar behaviors of the three models.  In each term, the first lower index represents the corresponding interaction parameter and, when present, the second lower index shows the variable on which the term depends. To distinguish the correlations, we use an upper index in the terms $A$ and $B$,  the value of which identifies the direction given by the vectors in Eq.~(\ref {Dvector}).
\begin{widetext}
\begin{center}
\line(1,0){450}
\end{center} 
\begin{eqnarray*}
B_{\alpha,\rho }&=& 
 \frac{1}{4}  \Delta \rho\\
B_{\alpha, C^j}&=&B_{\alpha, C^{j+1}}=B_{\alpha, C^{j+2}}=B_{\alpha,\rho, C^{j}}=B_{\alpha,\rho, C^{j+1}}=B_{\alpha,\rho, C^{j+2}}=0 
\end{eqnarray*}
\begin{center}
\line(1,0){250}
\end{center}
\begin{eqnarray*}
B_{\beta,\rho }&=& 
 \frac{2 }{3}  \left[(\nabla _1 \rho )^2+\nabla _2 \rho  \nabla _1
\rho +(\nabla _2 \rho )^2+\frac{3}{4} \rho  \Delta\rho \right] \\
 B_{\beta, C^0}&=&  \frac{1}{12}  (3  \Delta -4 \nabla _0^2) C^{0}\\ 
 B_{\beta, C^{1}}&=&  \frac{1}{12}  (3  \Delta -4 \nabla _1^2) C^{1}\\
B_{\beta, C^{2}}&=& \frac{1}{12}  (3 \Delta -4 \nabla _2^2)C^{2}\\
B_{\beta,\rho, C^0}&=&B_{\beta,\rho, C^{1}}=B_{\beta,\rho, C^{2}}=0
 \end{eqnarray*}
\begin{center}
\line(1,0){250}
\end{center}
\begin{eqnarray*}
 B_{\gamma,\rho }&=& 
 \frac{1}{3}  \left\{2 (5-7 \rho ) [(\nabla _1 \rho )^2+\nabla _2 \rho \nabla _1\rho+(\nabla _2 \rho )^2]
 +\frac{3}{4} (10-7 \rho ) \rho \Delta  \rho \right\} \\
B_{\gamma, C^0}&=&B_{\gamma, C^{1}}=B_{\gamma, C^{2}}=0\\
 B_{\gamma,\rho, C^0}&=& 
 \frac{1}{12} \left[(1-\rho ) (4 \nabla _0^2 +3 \Delta )C^{0} +  C^{0}(3\Delta  -14 \nabla _0^2)\rho  -18
    \nabla _0 \rho \nabla _0 C^{0}\right] \\
B_{\gamma,\rho, C^{1}}&=& 
 \frac{1}{12} \left[(1-\rho ) (4 \nabla _1^2 +3 \Delta )C^{1} +  C^{1}(3\Delta  -14 \nabla _1^2)\rho  -18
    \nabla _1 \rho \nabla _1 C^{1}\right] \\
 B_{\gamma,\rho, C^{2}}&=& 
 \frac{1}{12} \left[(1-\rho ) (4 \nabla _2^2 +3 \Delta )C^{2} +  C^{2}(3\Delta  -14 \nabla _2^2)\rho  -18
    \nabla _2 \rho \nabla _2 C^{2}\right] 
\end{eqnarray*}
\begin{center}
\line(1,0){450}
\end{center}
\begin{eqnarray*}
A_{\alpha }^j&=& 
 \frac{1}{3} (C^{j+1}+C^{j+2}-5 C^{j}) \\
 B_{\alpha, \rho }^j&=&-\frac{1}{6}  \left[ (\nabla _{j+1} +\nabla _{j+2} )\rho \right]^2 \\
B_{\alpha, C^{j}}^j&=&B_{\alpha, \rho, C^{j}}^j=B_{\alpha, \rho, C^{j+2}}^j=B_{\alpha, \rho, C^{j+2}}^j=0\\
B_{\alpha, C^{j+1}}^j&=& \frac{1}{24} \nabla _{j+2}^2 C^{j+1}\\
B_{\alpha, C^{j+2}}^j&=& \frac{1}{24} \nabla _{j+1}^2 C^{j+2}
\end{eqnarray*}
\begin{center}
\line(1,0){250}
\end{center}
\begin{eqnarray*}
A_{\beta }^j&=& \frac{1}{3} [C^{j+1}+C^{j+2}-2(1+3 \rho ) C^{j}] \\
 B_{\beta,\rho }^j&=& \frac{ 1}{6} \left[ (1-4\rho )(\nabla _{j} \rho)^2 -2 \rho(1-\rho)\nabla _{j+1}\nabla _{j+2} \rho\right] \\
 B_{\beta, C^{j}}^j&=&0\\
 B_{\beta, C^{j+1}}^j&=&\frac{1}{24} \nabla _{j+2}^2 C^{j+1} \\
 B_{\beta, C^{j+2}}^j&=&\frac{1}{24} \nabla _{j+1}^2 C^{j+2} \\
 B_{\beta,\rho, C^j}^j&=& 
 -\frac{1}{48} C^{j}  (28 \nabla _{j+1}\nabla _{j+2} +57
\Delta )\rho \\
B_{\beta,\rho, C^{j+1}}^j&=& 
 \frac{1}{3} [C^{j+1} \nabla_{j+1}-(\nabla _{j+2} C^{j+1})](\nabla _{j}+\nabla _{j+1}  )\rho \\
 B_{\beta,\rho, C^{j+2}}^j&=& 
 \frac{1}{3} [C^{j+2} \nabla_{j+2}-(\nabla _{j+1} C^{j+2})](\nabla _{j}+\nabla _{j+2}   )\rho
\end{eqnarray*}
\begin{center}
\line(1,0){250}
\end{center}
\begin{eqnarray*}
 A_{\gamma }^j&=& 
 \frac{1}{3} \left\{\rho  \left[-3 \rho  (1-\rho)^2 R^{-\xi} +(\rho +2)(C^{j+1} +C^{j+2})-(7 \rho +5)C^{j}  \right]\right\}-C^{j}  \\
 B_{\gamma,\rho }^j&=& 
 \frac{1}{12} \left\{\left(24 \rho ^2-26 \rho +3\right) (\nabla _{j} \rho)^2
+\rho(\rho -1) \left[ (3-14\rho )( \nabla _{j+1}^2 + \nabla_{j+2}^2 ) 
+6 (1-4 \rho) (\nabla _{j+1}\nabla_{j+2} )\right]\rho \right\}  \\
B_{\gamma, C^j}^j&=&B_{\gamma, C^{j+1}}^j=B_{\gamma, C^{j+2}}^j=0\\
 B_{\gamma,\rho, C^j}^j&=& 
 \frac{1}{24}  \left\{\vphantom{\frac{1}{4}} 4 (1-\rho) \rho  \nabla _{j}^2 C^{j}+4 (4 \rho -3) \nabla _{j} C^{j} \nabla _{j}\rho
\right.\\
& &\left.-C^{j}\left[146 (\nabla_{j}\rho)^2
-224\nabla _{j+1} \rho  \nabla _{j+2} \rho
+ (21\nabla_{j}^2-32\nabla_{j+1}\nabla_{j+2})\rho
+\rho  (158\nabla_{j}^2-128\nabla_{j+1}\nabla_{j+2})\rho
\right]\right\} \\
B_{\gamma,\rho, C^{j+1}}^j&=& 
\frac{1}{24}  \left\{\rho (\rho-1)(\nabla _{j+2}^2-8 \nabla _{j+1}\nabla_{j+2} -8\nabla _{j+1}^2 ) C^{j+1} +3\nabla _{j+2}^2C^{j+1}+
2C^{j+1}\nabla _{j}\rho (\nabla_{j}+16 \nabla _{j+2})\rho
\right.\\ 
& &\left.\qquad
+2(1+\rho) C^{j+1} (26 \nabla _{j+1}\nabla _{j+2} +5 \nabla_{j+1}^2 +9 \nabla _{j+2}^2 )\rho-64C^{j+1} \nabla _{j+1}\nabla _{j+2}\rho
\right.\\ 
& &\left. \qquad
+8 \nabla _{j+1} C^{j+1}[(2 \rho -1)\nabla _{j} +(\rho -1) \nabla _{j+2}]\rho -4 \nabla _{j+2} C^{j+1} [(\rho +1) \nabla _{j}+4 (2 \rho-1) \nabla _{j+2}]\rho\right\} \\
B_{\gamma,\rho, C^{j+2}}^j&=& 
\frac{1}{24}  \left\{\rho (\rho-1)(\nabla _{j+1}^2-8 \nabla _{j+2}\nabla_{j+1} -8\nabla _{j+2}^2 ) C^{j+2} +3\nabla _{j+1}^2C^{j+2}+
2C^{j+2}\nabla _{j}\rho (\nabla_{j}+16 \nabla _{j+1})\rho
\right.\\ 
& &\left.\qquad
+2(1+\rho) C^{j+2} (26 \nabla _{j+2}\nabla _{j+1} +5 \nabla_{j+2}^2 +9 \nabla _{j+1}^2 )\rho-64C^{j+2} \nabla _{j+2}\nabla _{j+1}\rho
\right.\\
& &\left. \qquad
+8 \nabla _{j+2} C^{j+2}[(2 \rho -1)\nabla _{j} +(\rho -1) \nabla _{j+1}]\rho -4 \nabla _{j+1} C^{j+2} [(\rho +1) \nabla _{j}+4 (2 \rho-1) \nabla _{j+1}]\rho\right\}
 \end{eqnarray*}
\begin{center}
\line(1,0){450}
\end{center} 
\end{widetext}

\bibliography{english}

\end{document}